\documentclass[11pt]{article}
\usepackage{authblk}
\usepackage{hyperref}
\usepackage{natbib}
\usepackage{graphicx}
\usepackage{verbatim}
\usepackage{placeins}

\usepackage[vscale=0.88,hscale=0.8,includehead,vmarginratio=1:9]{geometry}
\usepackage{times}

\begin{document}
	\title{Calibrating GONG Magnetograms with End-to-end Instrument Simulation III: Comparison, Calibration, and Results}
	\author[1]{Joseph Plowman}
	\author[2]{Thomas Berger}
	\affil[1]{National Solar Observatory, \url{jplowman@nso.edu}} 
	\affil[2]{University of Colorado at Boulder}
	\date{}
	\maketitle
	\begin{abstract}
		This is the last of three papers describing an `absolute' calibration of the GONG magnetograph using and end-to-end simulation of its measurement process. The simulation begins with a MURaM 3D MHD datacube and ends with a `synthetic magnetogram' of the corresponding magnetic field values as they would be observed by GONG. We determine a calibration by comparing the synthetic magnetic field measurements with the MURaM magnetic field values that produced them. The previous two papers have described the GONG measurement process (both instrument and data processing), our simulation of it, and the theory of magnetogram comparison and calibration. In this paper, we address some final points on calibration, combine all of this work into a set of calibration curves, and consider the results. We also review the results of the previous two papers for locality of reference. Our calibration indicates that GONG magnetograms underestimate weak flux by a factor of $\sim 2$ near disk center, but that factor decreases to $\sim 1$ as the line-of-sight approaches the limb. A preliminary investigation of the generalizability of these results suggests other instruments will be affected in a similar way. We also find that some differences in previous magnetograph comparisons are artifacts of instrumental resolution which do not reflect an intrinsic calibration difference, and the measurements are more similar than sometimes thought. These results are directly applicable to question of solar wind prediction model accuracies, particularly in the search for the cause of the common discrepancy between predicted solar wind magnetic flux at 1 AU and values measured {\em in situ} by current satellite missions.\\
		\noindent{\bf Keywords:} Instrumental~ Effects; Magnetic fields,~ Interplanetary; Magnetic fields,~ Models; Magnetic fields,~ Photosphere
	\end{abstract}

	\section{Introduction}\label{sec:intro}

	Measurement of the solar photospheric magnetic field is well-established and long-standing: routine observations have been now been made for over 75 years \citep[see][]{Babcock53, HowardBabcock60, HowardEtal83}. A variety of instruments currently make regular measurements of the photospheric magnetic field; this work concentrates on the `synoptic' measurements made by the GONG instruments, which are designed to make regularly recurring observations of the entire solar hemisphere visible from Earth. A primary use of these `synoptic magnetograms' is as the `boundary condition' for models of the coronal magnetic field. These in turn are the primary inputs to models that attempt to forecast the solar wind at Earth. Solar wind forecast models such as WSA/Enlil \citep[][used in operations by NOAA, the US Air Force (USAF), and the UK MetOffice]{ArgePizzo_JGR2000} provide the primary current predictions of geomagnetic storms due to the solar wind and Coronal Mass Ejections (CMEs). 

	One major issue with these measurements is that no magnetograms currently in use have been calibrated in any absolute sense, and a variety of apparent disagreements have between found of studies comparing these measurements made by different instruments. Further reinforcing the need for an absolute calibration of magnetograms, comparison of {\em in situ} measurements of magnetic flux in the solar wind at 1 AU (e.g, from the ACE and WIND spacecraft, \cite{1998StoneEtal_ACE_SSRv86_1S, 1995LeppingEtal_WindMFI_SSRv71_207L}) with model predictions based on photospheric magnetic field extrapolations into the solar wind are consistently at least two times higher than model predictions based on several different magnetogram inputs \citep{LinkerOpenFlux2017}. 

	We tackle this `absolute' calibration issue by developing an `end-to-end' simulation of the GONG measurement process: A MURaM 3D MHD simulation \citep[MURaM;][]{Rempel2015} provides a numerical `ground truth', which is used to produce a spectrum using Rybicki and Hummer (RH) radiative transfer model \citep{Uitenbroek_ApJ2001}. A GONG optical, polarimetric, and magnetogram processing model then simulates all of the major physical and numerical processes that comprise a full-disk magnetogram observation. By comparing these `synthetic magnetograms' with a `ground truth magnetogram' produced directly from the MURaM magnetic field, we produce a set of calibration curves for GONG magnetograms that can be used to correct both full-disk magnetograms and the synoptic maps that are created from them. 

	This is the last in a series of three papers on this work. In the first \citep{PlowmanEtal_2019I}, we gave an updated description of the GONG instrument and its magnetic field measurement process. In the second \citep{PlowmanEtal_2019II}, we considered the theory and goals of the calibration, and of magnetogram comparison in general. One major result of that paper is that, for each pixel, we calibrate to the entire spatial region sampled by that pixel rather than only within the `boundaries' implied by the physical pixel grid on the GONG CCDs. Thus, `absolute' calibration means that we compare the measurements to {\em all} of the ground truth values, combined, that have contributed to the measurement at that pixel, including those contributions due to the instrument PSF. We found that omission of the PSF from that calibration process lead to serious problems with the resulting calibration, in particular that it did not preserve flux relative to the ground truth.

	In this last paper \citep{PlowmanEtal_2019III}, we begin by addressing two remaining issues in the theory of calibration: the line-of-sight integration aspect of the `ground truth reduction' introduced in \cite{PlowmanEtal_2019II}, in Section \ref{sec:loseffects} and the actual fitting of the calibration curves, in Section \ref{sec:curvefitting}. We then proceed to unify these last two pieces of the analysis with those described in \cite{PlowmanEtal_2019I,PlowmanEtal_2019II}, producing the calibration curves for GONG. We then discuss the implications of these curves, make a preliminary check of their effects on the open flux, and review all of the conclusions of this work.


	The calibration problem can be summarized as follows:  Given the MURaM `ground truth', the simulated measurements made from them, and the real measurements, how do we produce a set of calibrated measurements corresponding to the real measurements? The ground truth consists of a very high resolution, {\em three}-dimensional magnetic field (or flux) datacube, whereas the measurements are a low-resolution, {\em two}-dimensional image. In our case with GONG and MURaM, of order $10^5$ ground truth values are integrated over to form each pixel's single magnetic field measurement. The calibrated measurements will also be a two-dimensional image with the same number of pixels as the measurements.

	In \cite{PlowmanEtal_2019II}, we specialized to the case where the calibrated flux measurement at a given pixel does not depends on the (uncalibrated) measurements only via their value {\em at} that pixel, not on the vallues at any of the {\em surrounding pixels}: for a uncalibrated measurement value, the calibrated measurement will always have the same value (some random scatter could be added as well, but this is no different than adding noise). This is necessary due to the limitations of potential multipixel methods and the small size of our calibration ground truth.  The calibration must therefore be a one dimensional function -- a calibration curve -- of the (uncalibrated) measured values.
	
\begin{figure}
		\begin{center}\includegraphics[width=\textwidth]{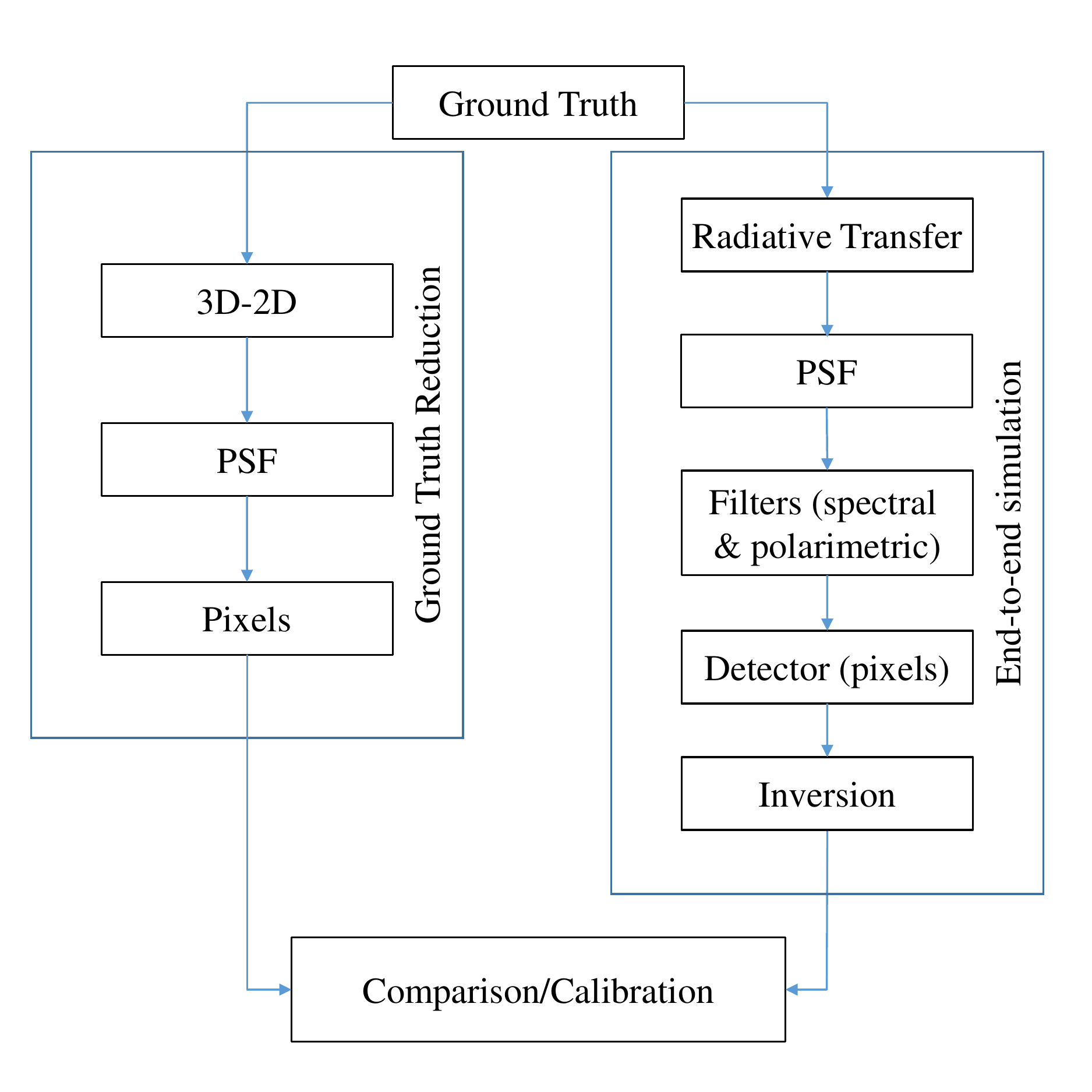}\end{center}
		\caption{Flow chart of the calibration process. On the right is the end-to-end simulation of the measurement process, while on the left is the `ground truth reduction' which places the ground truth in a form that can be directly compared with the measurements. This is necessary because some parts of the measurement process are irreversible.}\label{fig:calibration_boxdiagram}
	\end{figure}

	We determine these calibration curves by fitting them to `pixel-to-pixel' scatter plots obtained from the ground truth and corresponding synthetic measurements: The scatterplot has one point for each pixel in the synthetic measurement (i.e., the magnetogram image). The measured values are on the x axis, while the y axis is what would be obtained if the measurements were perfectly calibrated, which are produced from the ground truth. The overall calibration process is shown in Figure \ref{fig:calibration_boxdiagram}. We call the process of producing the `perfectly calibrated' values `ground truth reduction'. The ground truth reduction has two components; 3-D to 2-D (i.e., along the line of sight), and high resolution to low resolution (i.e., in the plane of the sky).
	
	\cite{PlowmanEtal_2019II} concentrated on the plane-of-sky component of the ground truth reduction. We demonstrated that flux conservation is an essential metric for any such calibration, both because the measurements (being based on area-integrated quantities) are themselves akin to fluxes, and because a calibration that does not conserve flux will give incorrect 3D magnetic field extrapolations even in the simplest case (i.e., a potential field). We then showed that a calibration that conserved flux was obtained from a scatter plot (or point cloud) of the synthetic ground truth against the synthetic measurements, but only if the ground truth reduction resampled the ground truth to the pixel scale of the measurements {\em and} applied the instrument PSF. Omitting the PSF resulted in a calibration that did not conserve flux. We concluded that the `perfectly calibrated' measurements must include the instrument PSF. We argued that this is not only necessary, but fitting: the measurements are area-integrated quantities, like fluxes, and much of the `weight' of the area integration comes from outside the pixel boundary. For GONG, only about $10\%$ of the weight comes from inside the pixel boundary, so it is measuring more flux `outside' the pixel more than `inside' it. It is therefore unsurprising that attempting to construct a calibration while ignoring most of the flux contributing to the measurement would be unsuccessful.
	
	\section{Line of sight integration effects: From 3D ground truth to 2D}\label{sec:loseffects}
	
	We now turn to the line-of-sight aspects of the ground truth reduction. Line of sight integration effects are considerably more difficult to treat systematically than the spatial (plane-of-sky) resolution effects. Rather than having a PSF that applies in a consistent way across all pixels, there is instead a contribution function of height for each wavelength which is different for every line of sight. Each of these contribution functions depends on the solar plasma parameters, including the magnetic field. The dependence is such that two wavelengths along the same line of sight may be weighted toward different regions of the plasma, with different magnetic fields, even if both are near the line center. These differences become more dramatic as the inclination angle moves away from the vertical (i.e., as latitude and/or longitude move away from disk center). 

\begin{figure}
		\begin{center}
			\includegraphics[width=0.19\textwidth]{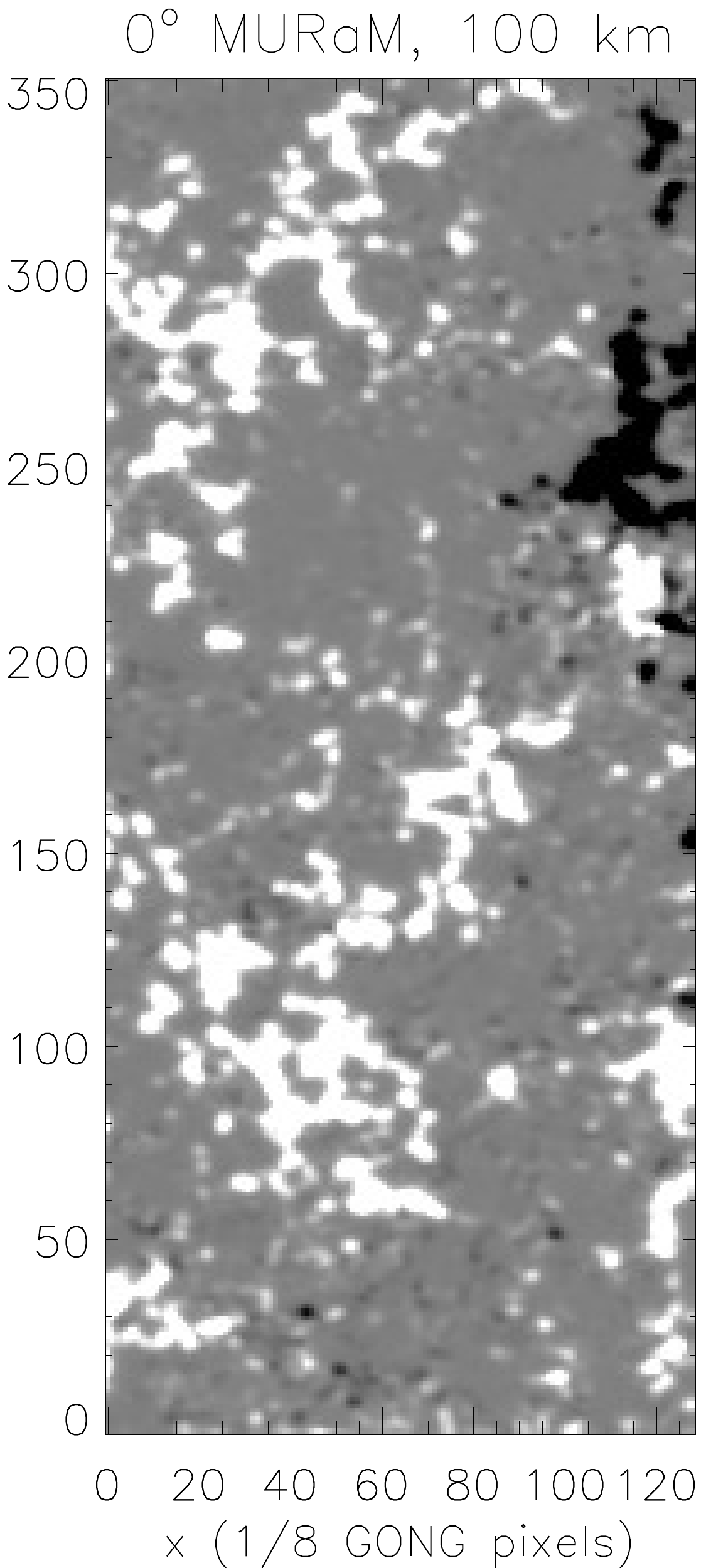}
			\includegraphics[width=0.19\textwidth]{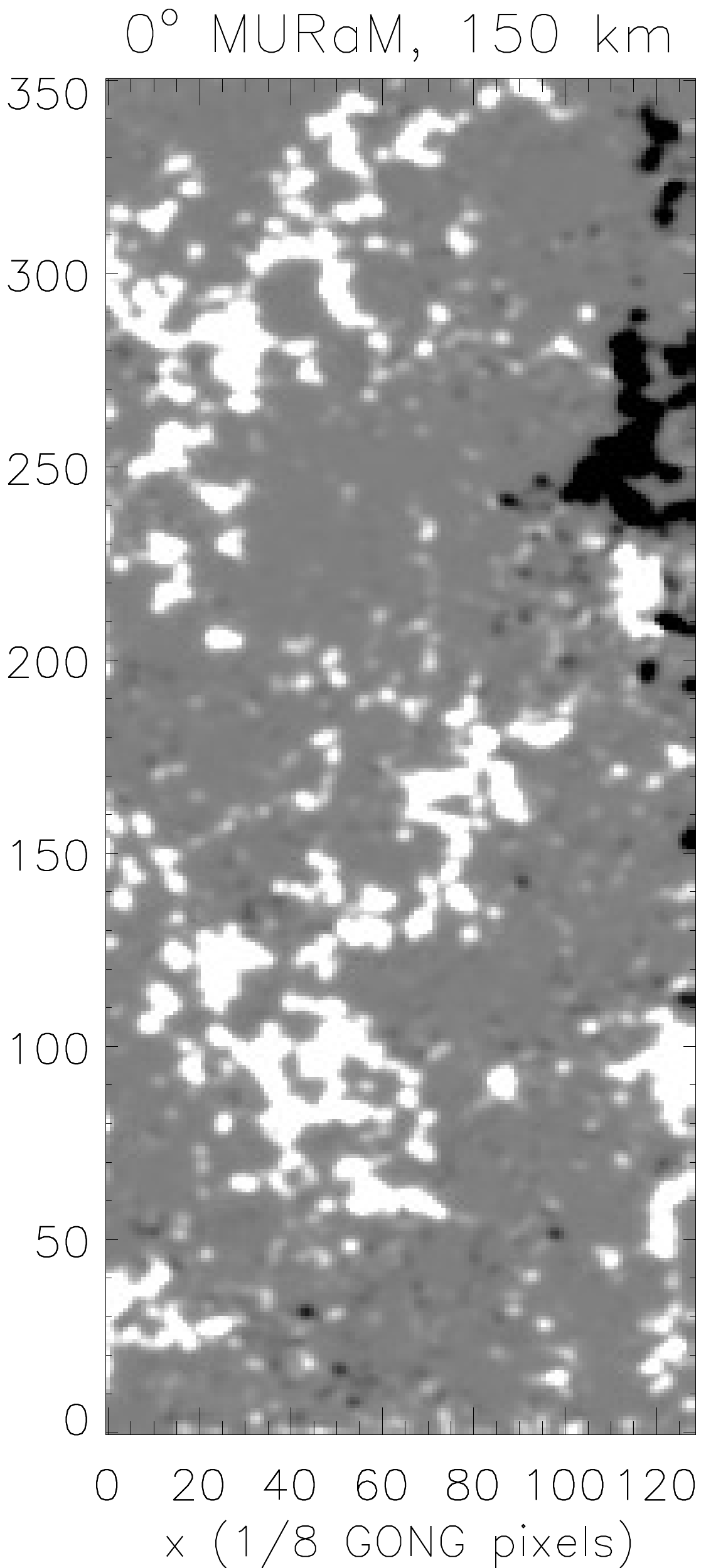}
			\includegraphics[width=0.19\textwidth]{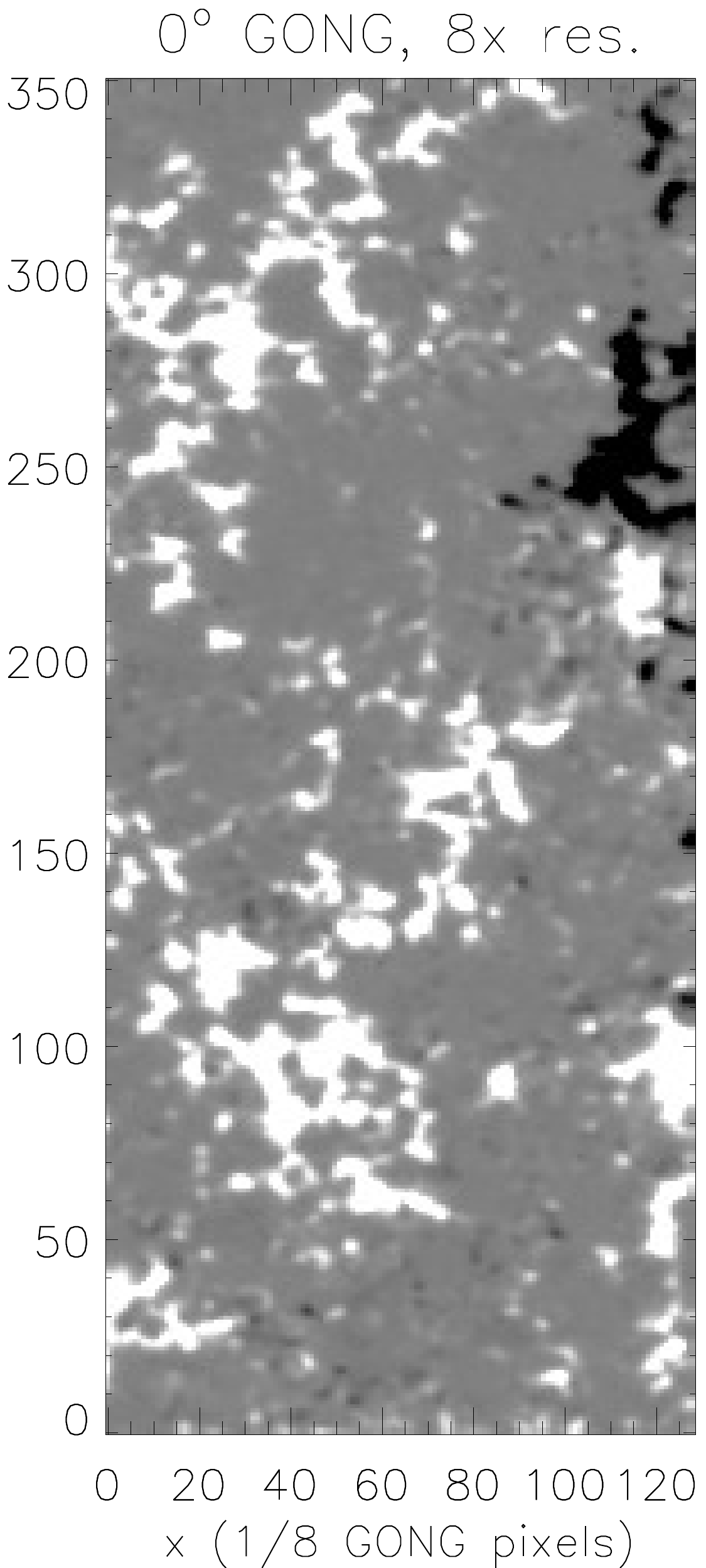}
			\includegraphics[width=0.19\textwidth]{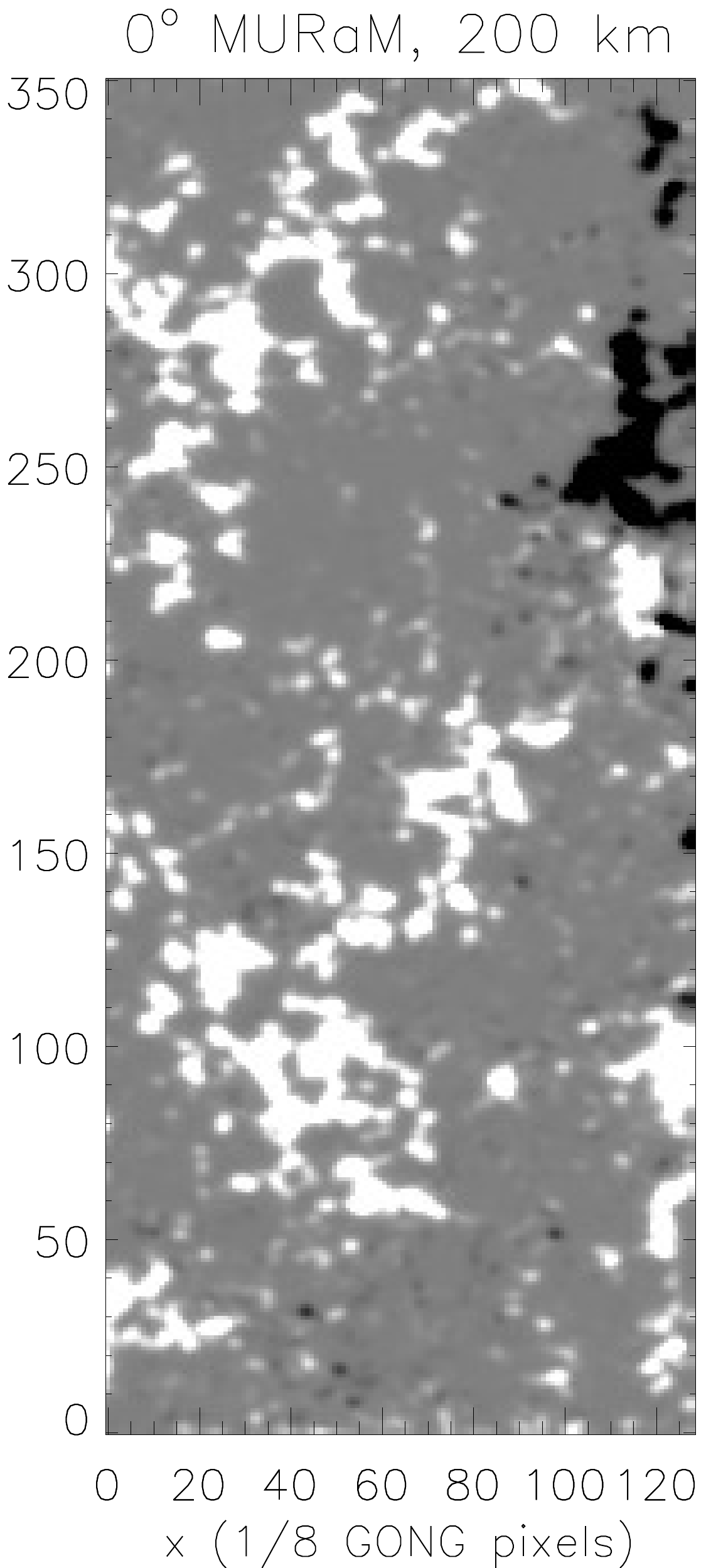}
			\includegraphics[width=0.19\textwidth]{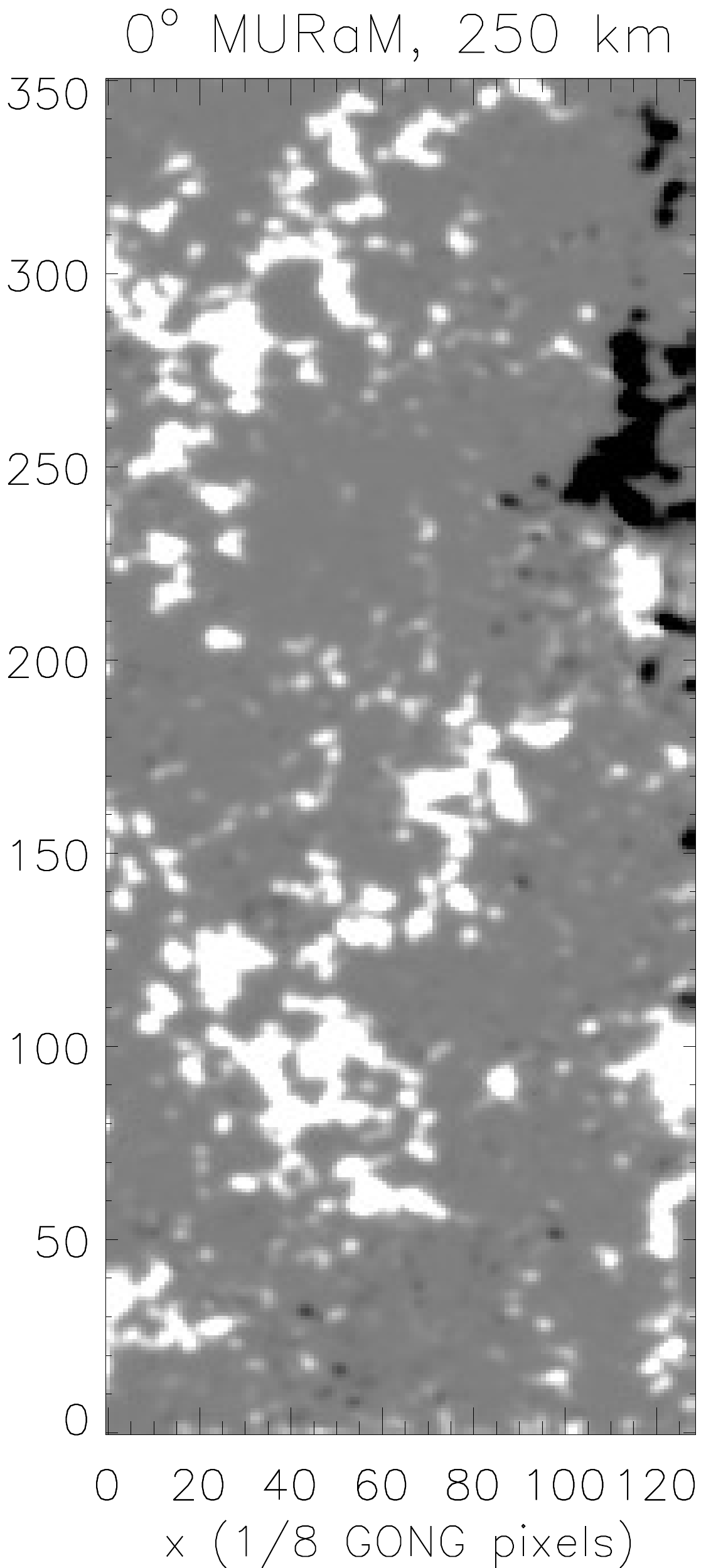}
		\end{center}
		\caption{Comparison of line-of-sight field strengths from the MURaM `ground truth' field (left and right images) with magnetogram from GONG simulator (center image), at 8 times GONG's native resolution and with vertical viewing angle. The MURaM ground truth are taken at heights of 150 to 250 km above the height of $\tau=1$ in the continuum (also with vertical viewing angle). Variation of the vertical MURaM field with height is small, and the GONG simulator magnetogram is most similar to the 150-200 km height range.}\label{fig:vertical_field_correspondence_images}
	\end{figure}

\begin{figure}
		\begin{center}
			\includegraphics[width=0.19\textwidth]{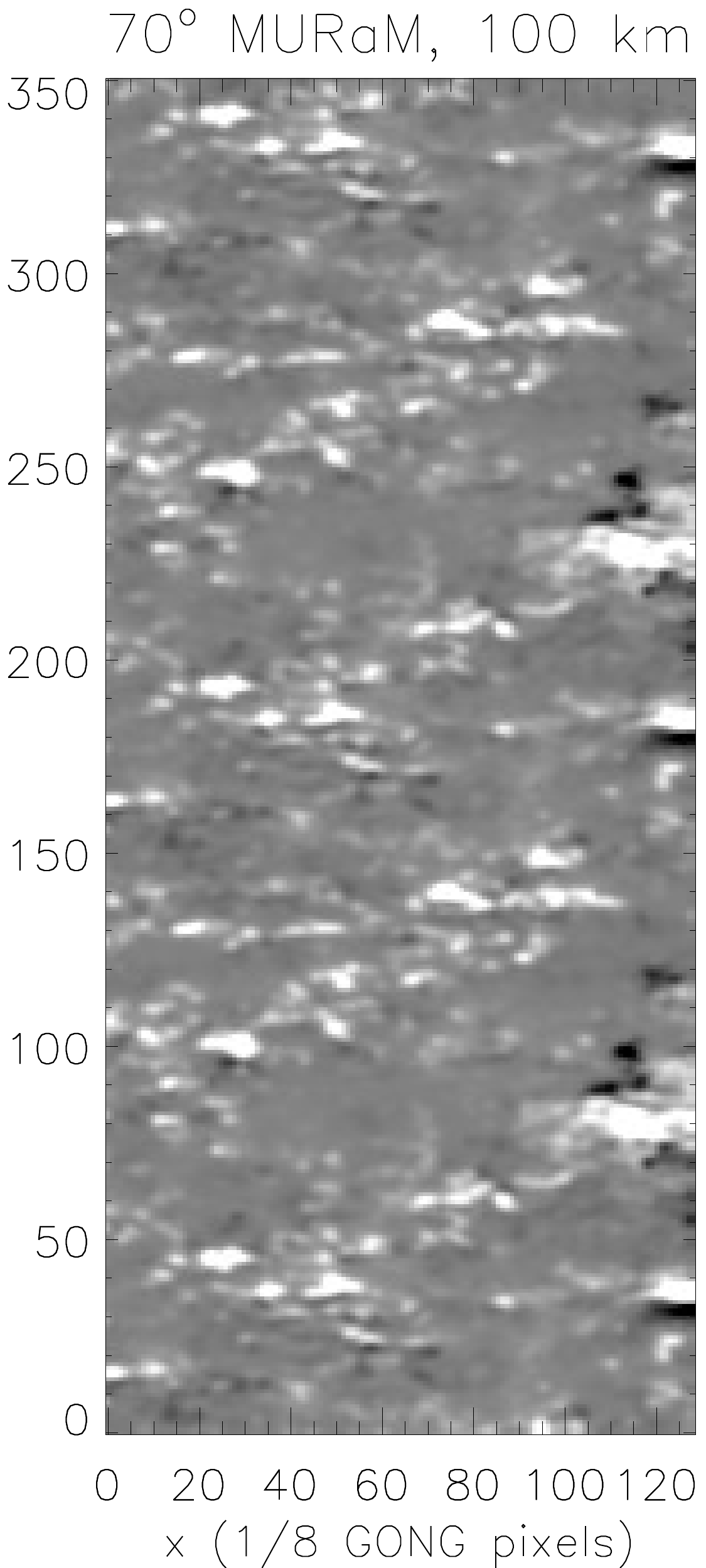}
			\includegraphics[width=0.19\textwidth]{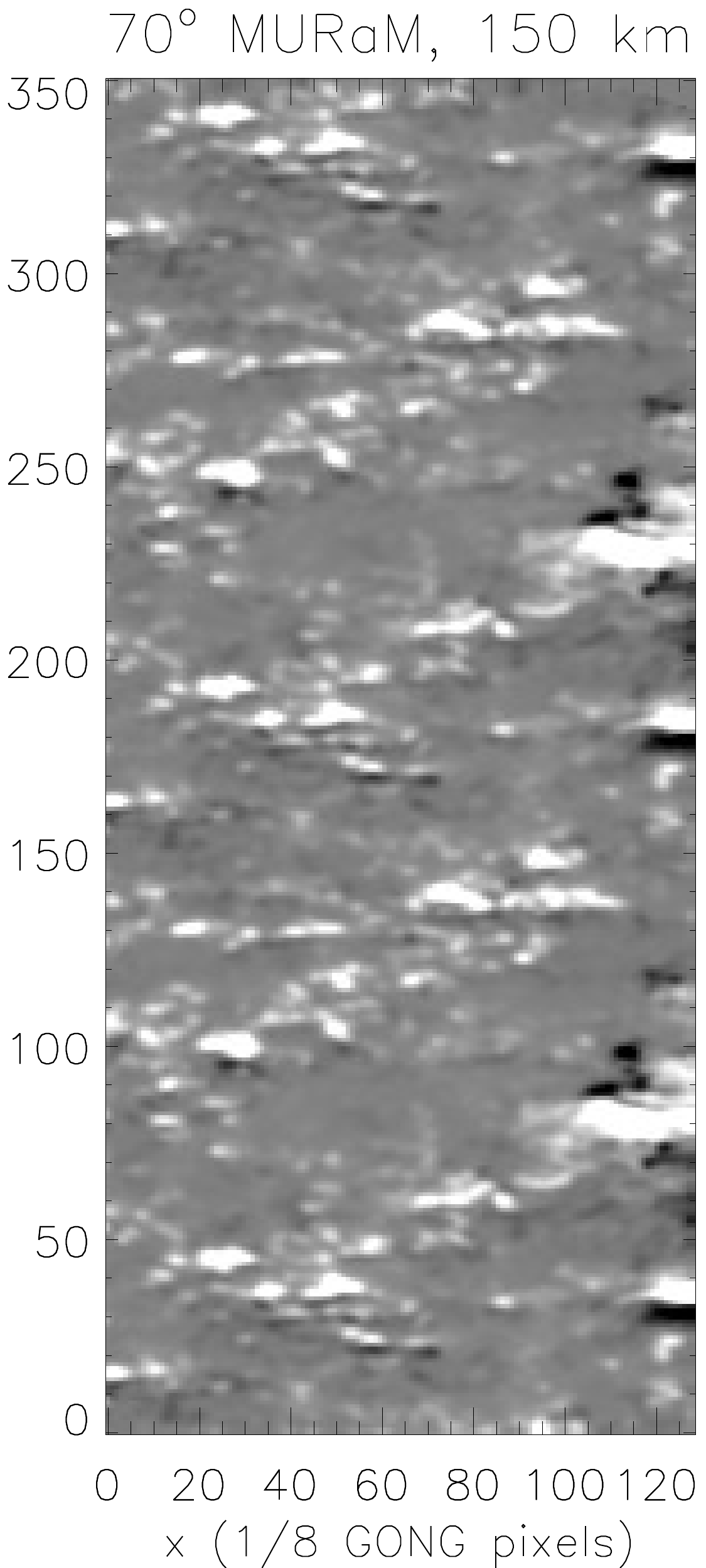}
			\includegraphics[width=0.19\textwidth]{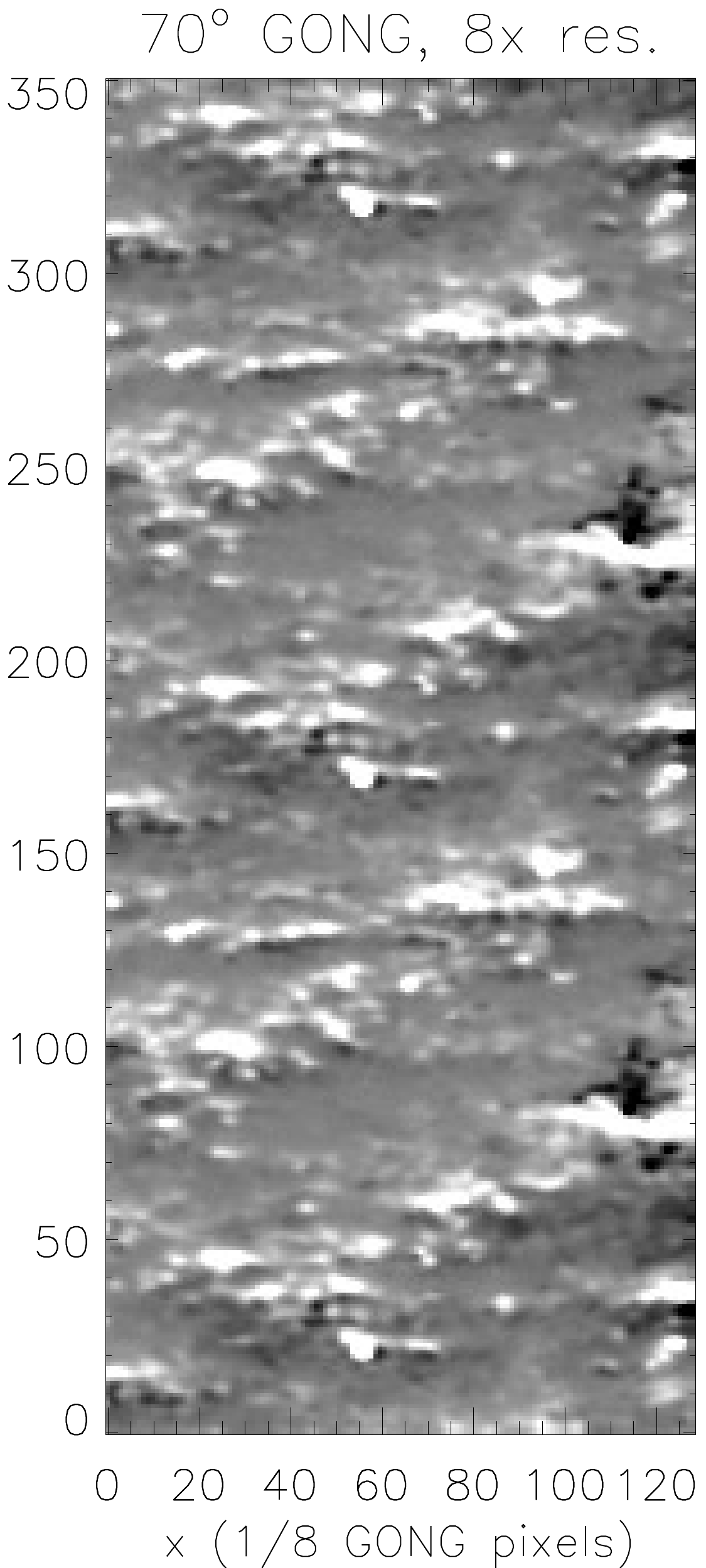}
			\includegraphics[width=0.19\textwidth]{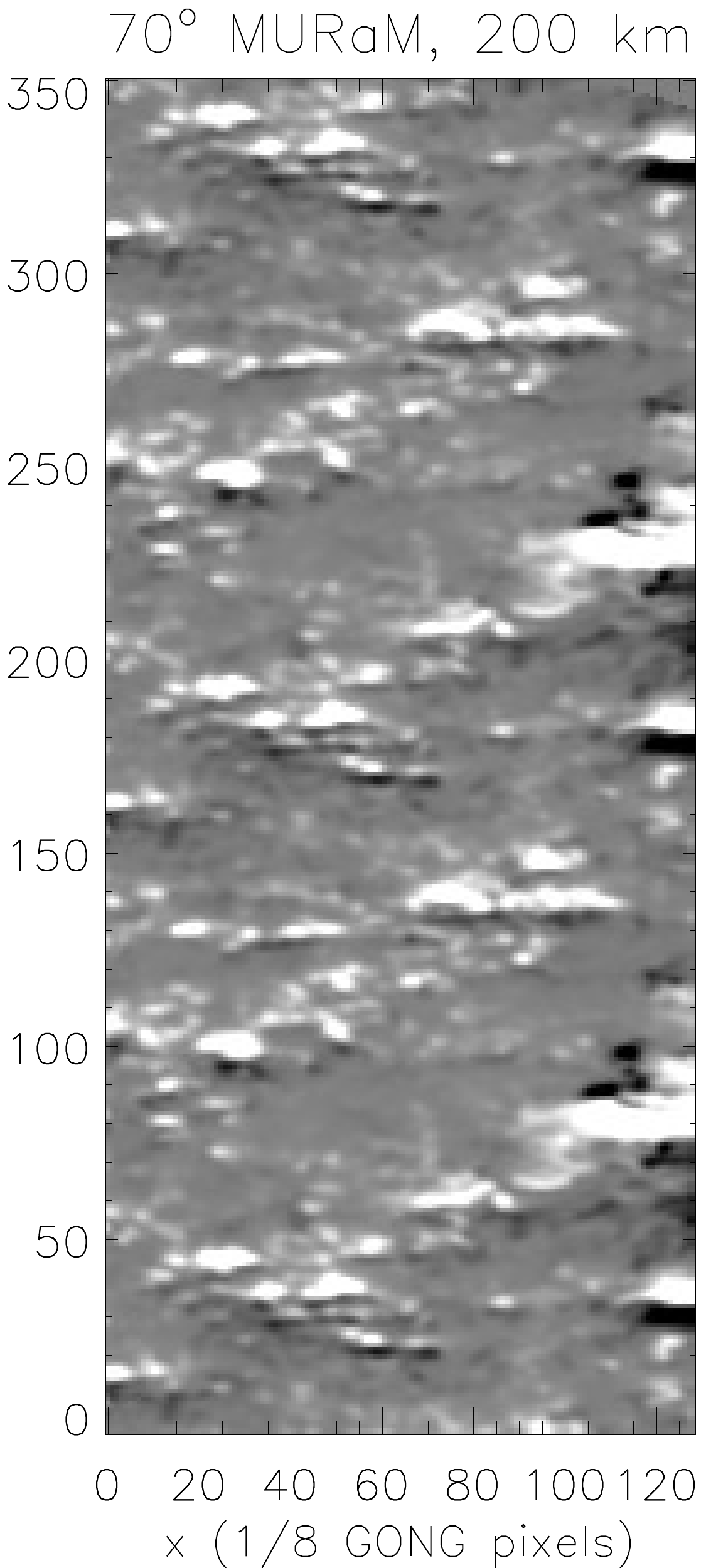}
			\includegraphics[width=0.19\textwidth]{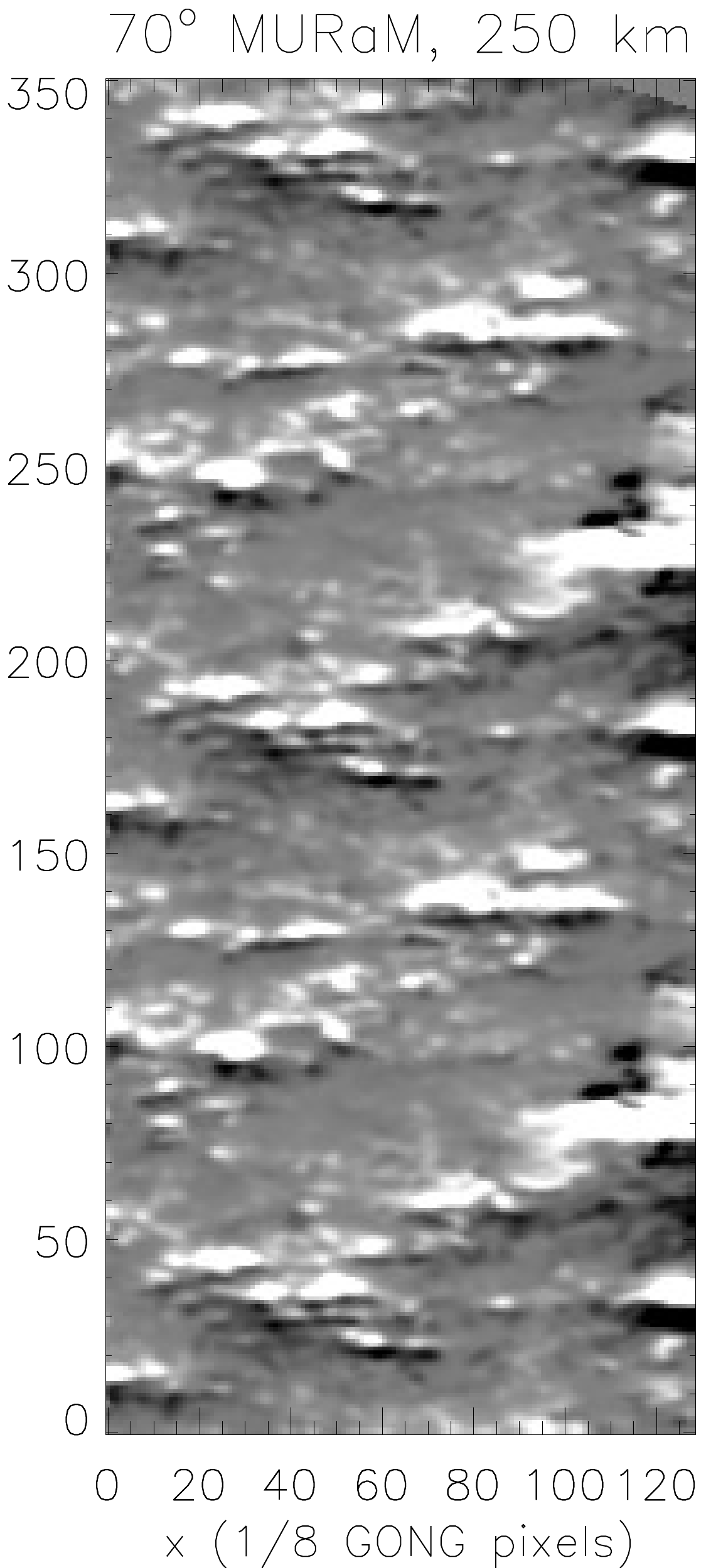}
		\end{center}
		\caption{Comparison of line-of-sight field strengths from the MURaM `ground truth' field (left and right images) with magnetogram from GONG simulator (center image) at 8 times GONG's native resolution and with 70 degree viewing angle. The MURaM ground truth are taken at heights of 150 to 250 km above the height of $\tau=1$ in the continuum (also with 70 degree viewing angle). Variation of this inclined MURaM field with height is significant (in part because the variation in the continuum $\tau=1$ height is more significant at 70 degrees), but the GONG simulator magnetogram is still most similar to the 150-200 km height range.}\label{fig:70degree_field_correspondence_images}
	\end{figure}

	As a result, there is no specific height for which the magnetic field along the line-of-sight direction exactly corresponds to the field value inferred from the Zeeman signal, neither is there a single weighting function which represents the effects of line-of-sight integration on the magnetic field. For the vertical, the field values can be fairly close (Figure \ref{fig:vertical_field_correspondence_images}), but the correspondence gets progressively worse as the inclination angle increases (Figure \ref{fig:70degree_field_correspondence_images} shows the 70 degree case).

	Because there is no such specific height, and the contribution functions vary considerably in height and wavelength, we have instead chosen to average over a range of heights: from 100 to 250 km above the height of continuum $\tau=1$. This reflects the range of typical contribution function heights -- evidenced by figures \ref{fig:vertical_field_correspondence_images} and \ref{fig:70degree_field_correspondence_images}, which suggest that the ground truth is most similar to the measurement at around 175 km above $\tau=1$. The range we chose is based on this consideration. We have checked a variety of such choices and the dependence on exact choice of height and averaging range is not strong. We defer showing examples of the scatter plots produced by this ground truth reduction until the results in Section \ref{sec:results}.


	\subsection{Calibration curve fitting}\label{sec:curvefitting}


	In the appendices of \citep{PlowmanEtal_2019II}, we mentioned two `per-pixel' curve fitting methods. One (histogram equating) does not technically use a scatter plot at all, but simply matches the one-dimensional distributions of per-pixel fluxes of the two axes. We showed that it still requires the same ground truth reduction (including PSF), even though it does not use the scatterplots directly and makes no use of an explicit correspondence between the two data sets being compared. The other method fits a curve to the per-pixel scatter plot, and therefore makes use of direct correspondence between pixels of the measurement and of the `reduced' ground truth. We prefer to base our calibration on the more direct fitting method, where the correspondence is made explicit. The method described here is equivalent to the latter bin-wise `flux-conserving' method described in \cite{PlowmanEtal_2019II} in the limit of small bin size and large numbers of points in each bin, but also ensures bin-wise flux conservation for large bins and smaller numbers of points in each bin. 


	We begin, as before, by dividing the point clouds into bins (using $x_i$ to denote the bin boundaries) in the {\em measured} values.
	The calibration curve is taken to be a linear interpolated function, with the nodes of the linear interpolation at the centers of each bin. In terms of the node values $y_i$ and bin {\em centers} $x_i'$ (located at $(x_i+x_{i+1})/2$), the calibration for a given measured field $m_{kl}$ is given by the usual linear interpolation formula,
	\begin{equation}
		c_{kl} = \frac{y_{i+1}(m_{kl}-x_i')-y_i(x_{i+1}'-m_{kl})}{x_{i+1}'-x_i'},
	\end{equation}
	where the index $i$ is chosen so that $m_{kl}$ falls between $x_i'$ and $x_{i+1}'$, or chosen to extrapolate if $m_{kl}$ falls outside the bin center range: If $m_{kl} < x_0'$ (measurement to be calibrated is less than the lowest bin center), we use $i=0$, and if $m_{kl}>x_n'$ (measurement to be calibrated is greater than the highest bin center), we use $i=n$. To determine $y_i$ (the calibration values at the node points) we require that, for the ground truth and GONG simulator data used to produce the calibration, the calibrated net flux in each bin must equal the net flux from the ground truth in the same bin:
	\begin{equation}\label{eq:linear_interp_binfluxconservation}
		\sum_{x_i < m_{kl} < x_{i+1}} c_{kl} = \sum_{x_i < m_{kl} < x_{i+1}} a_{kl}
	\end{equation}
	Here the summation is over all points whose $m_{kl}$ falls in the $i$th bin (i.e., between $x_i$ and $x_{i+1}$). The bins cover the entire range of the $m_{kl}$, with the first and last bin boundaries chosen so that they each contain 1600 points to reduce errors. The remaining, inner, bins are uniformly distributed between the upper bound of the lower bin and the lower bound of the upper bin. With the node points located at the bin centers and the edge treatment described above, this results in a linear tridiagonal system which can be solved for the $y_i$ by the usual matrix methods. The calibration curve fit can then be applied to a magnetogram (measured values $m_{ij}$) by linear interpolation of the curve $y_i(x_{i}')$ at each $m_{ij}$ -- i.e., the `$x$' axis values at which the interpolated values of the curve are to be computed are the $m_{ij}$.

	Because the MHD simulation has a significant flux bias, especially when the sunspot periphery is carefully cropped out, we also `mirror' the point cloud to fill in the underrepresented negative polarity. That is, given a set of `measured' and `ground truth' values $m_{ij}$ and $a_{ij}$, we append $-m_{ij}$ and $-a_{ij}$ to the set. We make separate scatter plots for the the `quiet sun' (non-sunspot) pixels and for the sunspot pixels. Sunspot pixels are identified as those which are over 31 Mm from the sunspot center (a conservative cut to avoid contamination from the sunspot).

	\section{GONG Simulator Calibration and Results}\label{sec:results}

	Descriptions of the MURaM simulation and our simulation of the GONG instrument can be found in \cite{PlowmanEtal_2019I}, and we will not attempt to reiterate them here. However, we will quote the following paragraph for reference:
	
	\begin{quotation}
		These spectral cubes are produced from MURaM simulations by the RH radiative transfer code. The MURaM simulations are much smaller than the whole sun due to computational limitations, so we tile them at several scales to cover the GONG detector plane. The intention is to increase the number of points in the calibration (via the tiling), and have some ability to investigate the effect of instrument resolution built into the simulator results. The effect of different viewing angles are investigated with an independent run of the simulator (resulting in a separate image), rather than tackling the thorny problem of stitching the simulation over a sphere. The tiling, and multiple resolutions, and separate images for each viewing angle are illustrated in Figure 11.
	\end{quotation}

\begin{figure}
		\begin{center}\includegraphics[width=\textwidth]{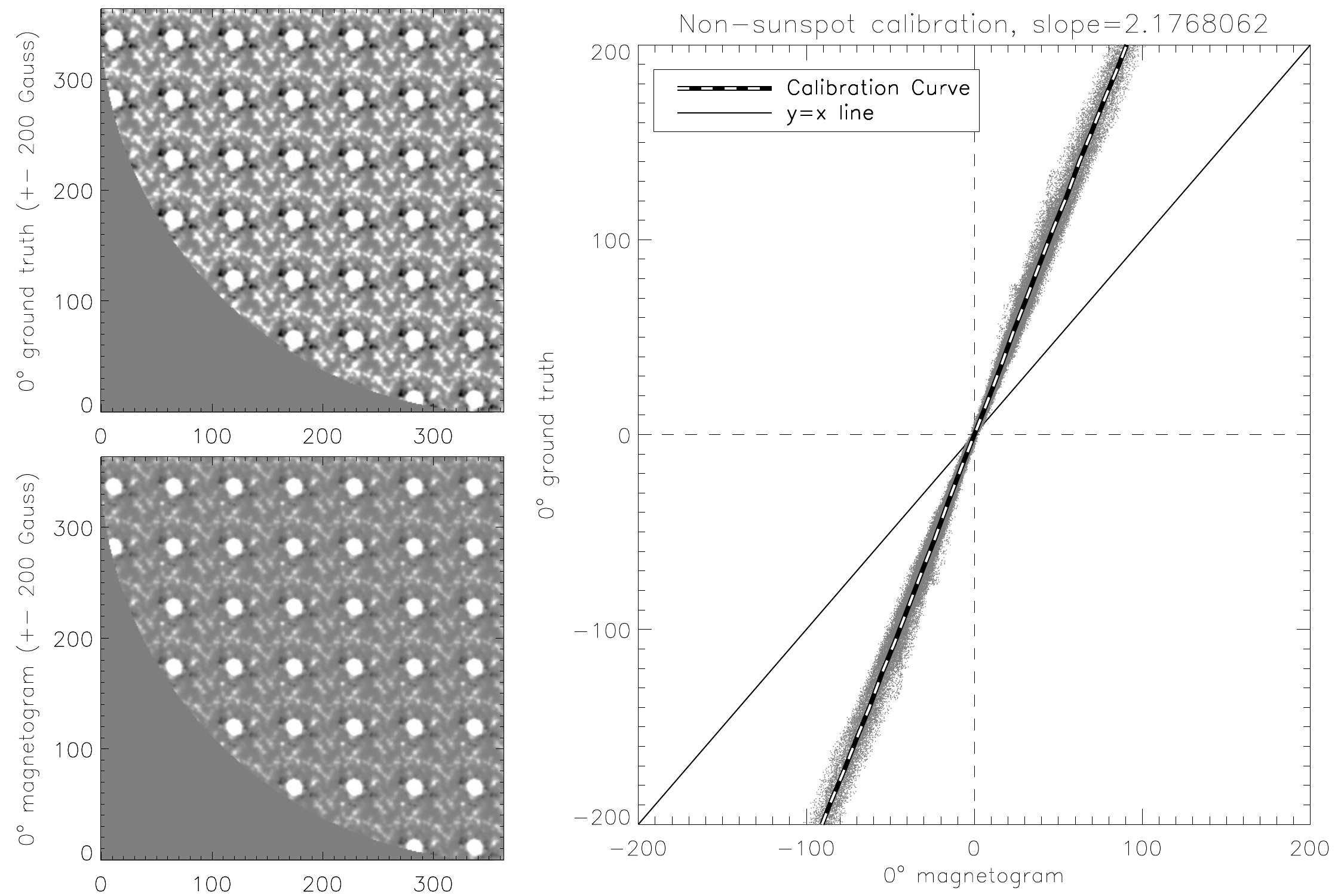}\end{center}
		\caption{Calibration and scatter plot for 0 degree (vertical, disk center) inclination. Left shows ground truth magnetogram, center shows GONG simulator magnetogram, right shows point cloud and calibration curve.}\label{fig:example_scatterplot}
	\end{figure}

\begin{figure}
		\begin{center}\includegraphics[width=\textwidth]{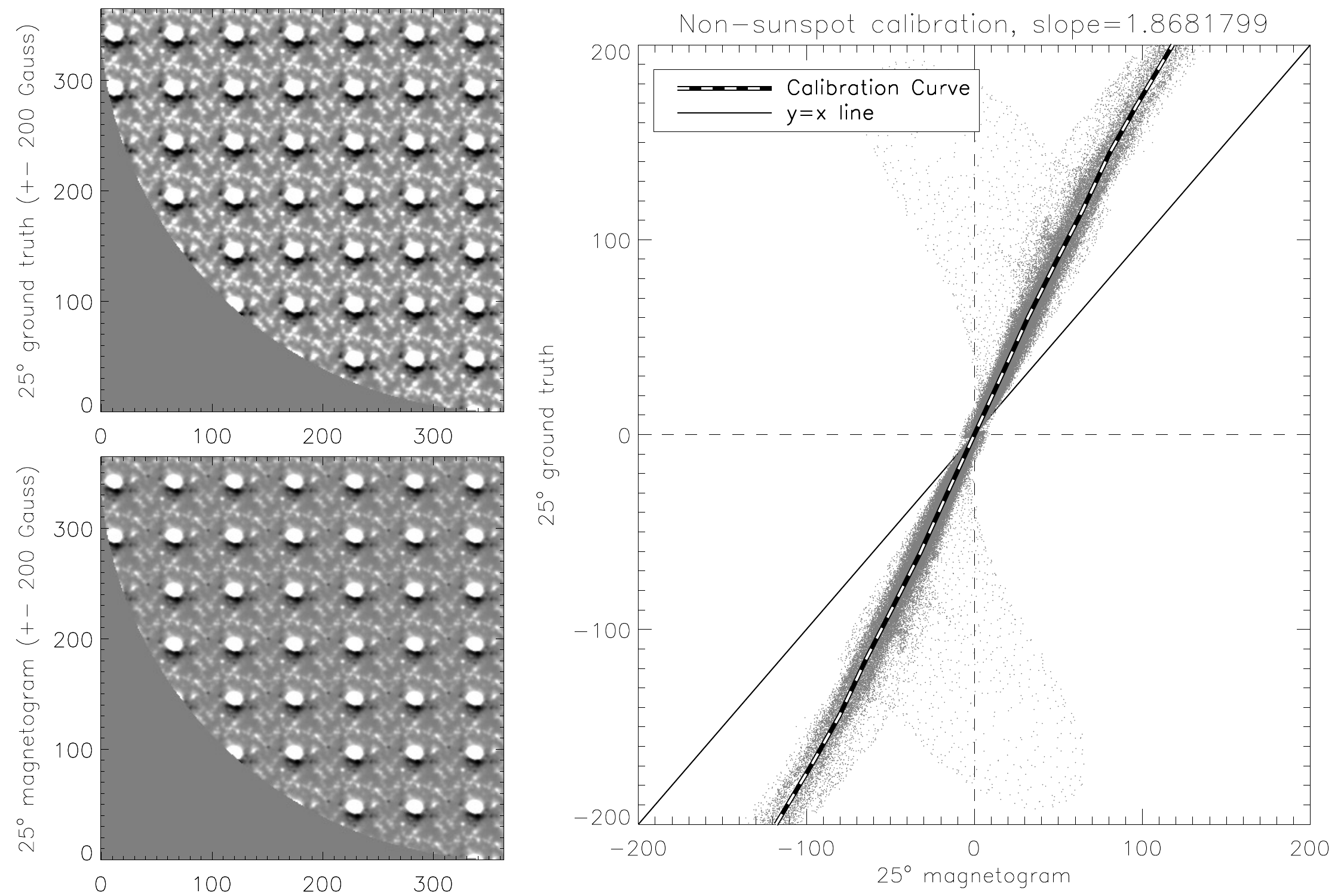}\end{center}
		\caption{Calibration and scatter plot for 25 degree inclination. Left shows ground truth magnetogram, center shows GONG simulator magnetogram, right shows point cloud and calibration curve.}\label{fig:cal_curve_25deg}
	\end{figure}

\begin{figure}
		\begin{center}\includegraphics[width=\textwidth]{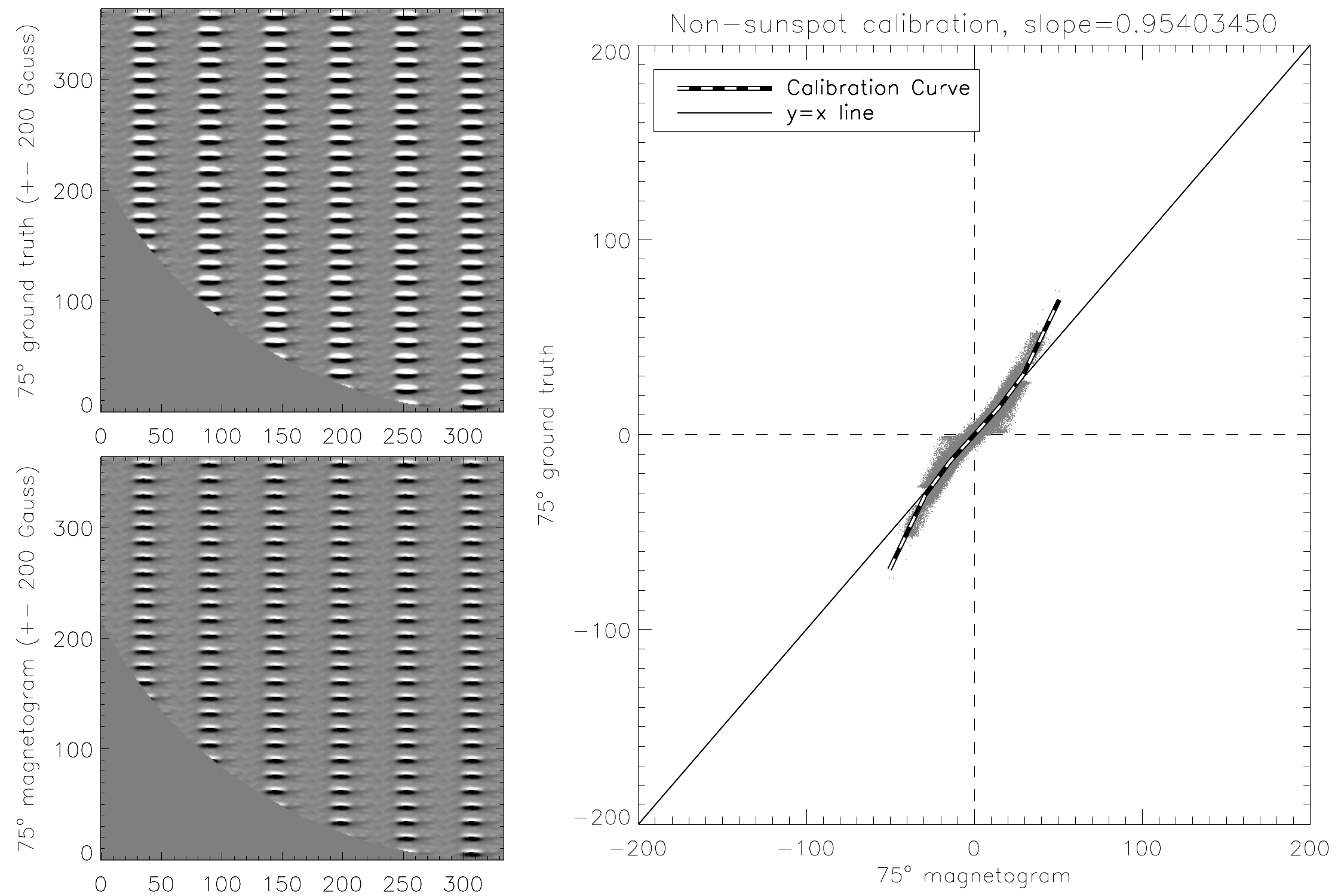}\end{center}
		\caption{Calibration curve for 75 degree inclination. Left shows ground truth magnetogram, center shows GONG simulator magnetogram, right shows point cloud and calibration curve.}\label{fig:cal_curve_75deg}
	\end{figure}

\begin{figure}
		\begin{center}\includegraphics[height=0.4\textheight]{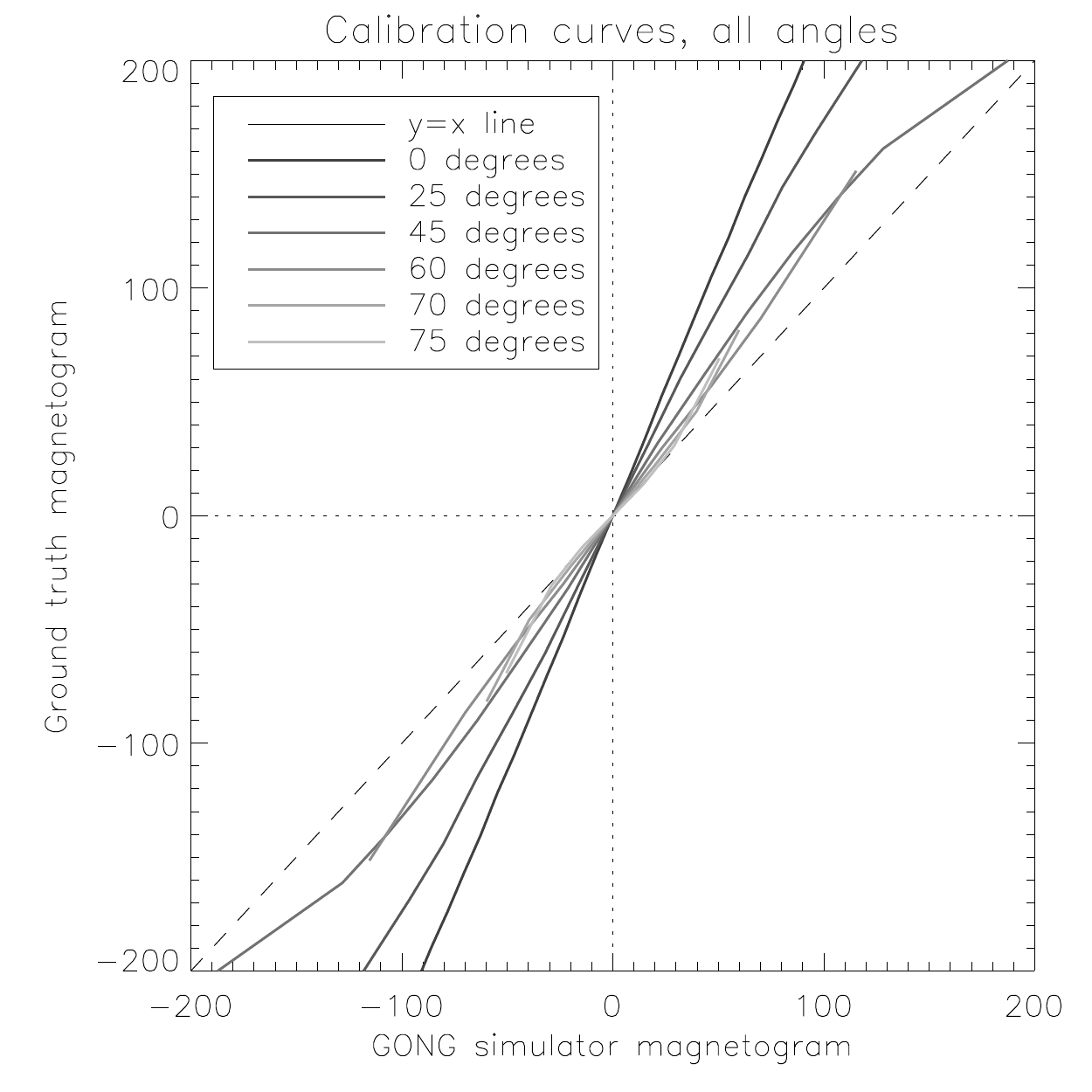}\end{center}
		\caption{Calibration curves for all six inclinations/latitudes considered.}\label{fig:cal_curves_combined}
	\end{figure}

	We will first consider the `native GONG' resolution quadrant (lower left) of the results, and the `quiet sun' pixels, which we identify as those which are over 31 Mm from the center of the sunspot (a conservative cut to avoid contamination). This is the most interesting resolution for GONG calibration purposes, and the non-sunspot regions are more important for space weather. Figures \ref{fig:example_scatterplot}, \ref{fig:cal_curve_25deg}, and \ref{fig:cal_curve_75deg} show the results of our calibration procedure for 0, 25, and 75 degree latitude. Figure \ref{fig:cal_curves_combined} shows curves for all 6 latitudes (0, 25, 45, 60, 70, and 75 degrees) together on the same graph.

	We find a `quiet sun' (i.e., non-sunspot) calibration factor of $\sim 2.2$ at disk center, which drops to near one at high latitudes. This is very similar to the behavior seen for the nonlinear `convective blueshift' effects alone in \cite{PlowmanEtal_2019II}, so those effects are likely to dominate the calibration factors at disk center. At higher latitudes, the quiet sun calibration factor does not drop as quickly: higher-latitude calibration factors for the full simulation are slightly higher than for the nonlinearity effects in \cite{PlowmanEtal_2019II}. Evidently, other effects become more significant at higher latitudes. 

	Particularly interesting is the upturn in the calibration factor for strong flux/field pixels at higher latitudes -- although there is not a significant miscalibration of the {\em weakest} high-latitude fields, {\em stronger} ($\sim 30$ Gauss) high-latitude fields are underestimated. Moreover, the factor by which they are underestimated increases with latitude. This result should be considered preliminary, and extrapolation to even higher latitude with our relatively small simulation volume is fraught. However, this effect may provide an additional boost to the calibration factor for polar fields. Comparison of magnetograms as regions rotate onto the limb may prove informative for this question, albeit complicated by the time variations of those observations and the physical differences between polar and near-equatorial solar regions. Simultaneous observations from two widely separated perspectives (e.g., one from Earth and one significantly away from the Earth-Sun line) are likely to shed considerable light onto this issue.

	The other three quadrants of the synthetic GONG images allow us to investigate the effect of spatial resolution on the calibration -- i.e., how does it change if GONG were 2, 4, or 8 times its present resolution? In the interests of brevity, the results are summarized in tabular form in Table \ref{tab:calfactors}. The point cloud relationships are similar to those at GONG's native resolution, except there is more scatter at high resolution, and the calibration factors gradually become smaller as the resolution increases. An important point of comparison is between GONG and HMI, whose resolution is nearly eight times that of GONG, including atmospheric seeing. Table \ref{tab:calfactors} therefore indicates that the resolution difference results in a relative calibration factor of $\sim 1.2$ between GONG and HMI for weak-field non-sunspot regions at disk center, before any other instrument differences are taken into account. However, this relative difference drops as the viewing angle increases, resulting in very similar factors at all resolutions near the limb. Instrumental differences other than resolution alone (different spectral lines, spectral resolution, etc) could easily change the comparison, of course. 

	\begin{table}[h]
		\begin{center}\begin{tabular}{|l|c|c|c|c|c|c|}
			\hline
			Resolution & $0^\circ $ & $25^\circ $ & $45^\circ $ & $60^\circ $ & $70^\circ $ & $75^\circ $ \\
			\hline
			\hline
			1x GONG resolution & 2.17 & 1.89 & 1.59 & 1.34 & 1.15 & 0.98 \\
			2x GONG resolution & 2.10 & 1.81 & 1.44 & 1.24 & 1.12 & 1.01 \\
			4x GONG resolution & 1.96 & 1.66 & 1.24 & 1.08 & 1.02 & 0.98 \\
			8x GONG resolution & 1.59 & 1.49 & 1.13 & 1.02 & 0.96 & 0.89 \\
			\hline
		\end{tabular}\end{center}
		\caption{Weak-field non-sunspot calibration factors as a function of resolution and viewing angle.}\label{tab:calfactors}
	\end{table}

\begin{figure}
		\begin{center}\includegraphics[width=\textwidth]{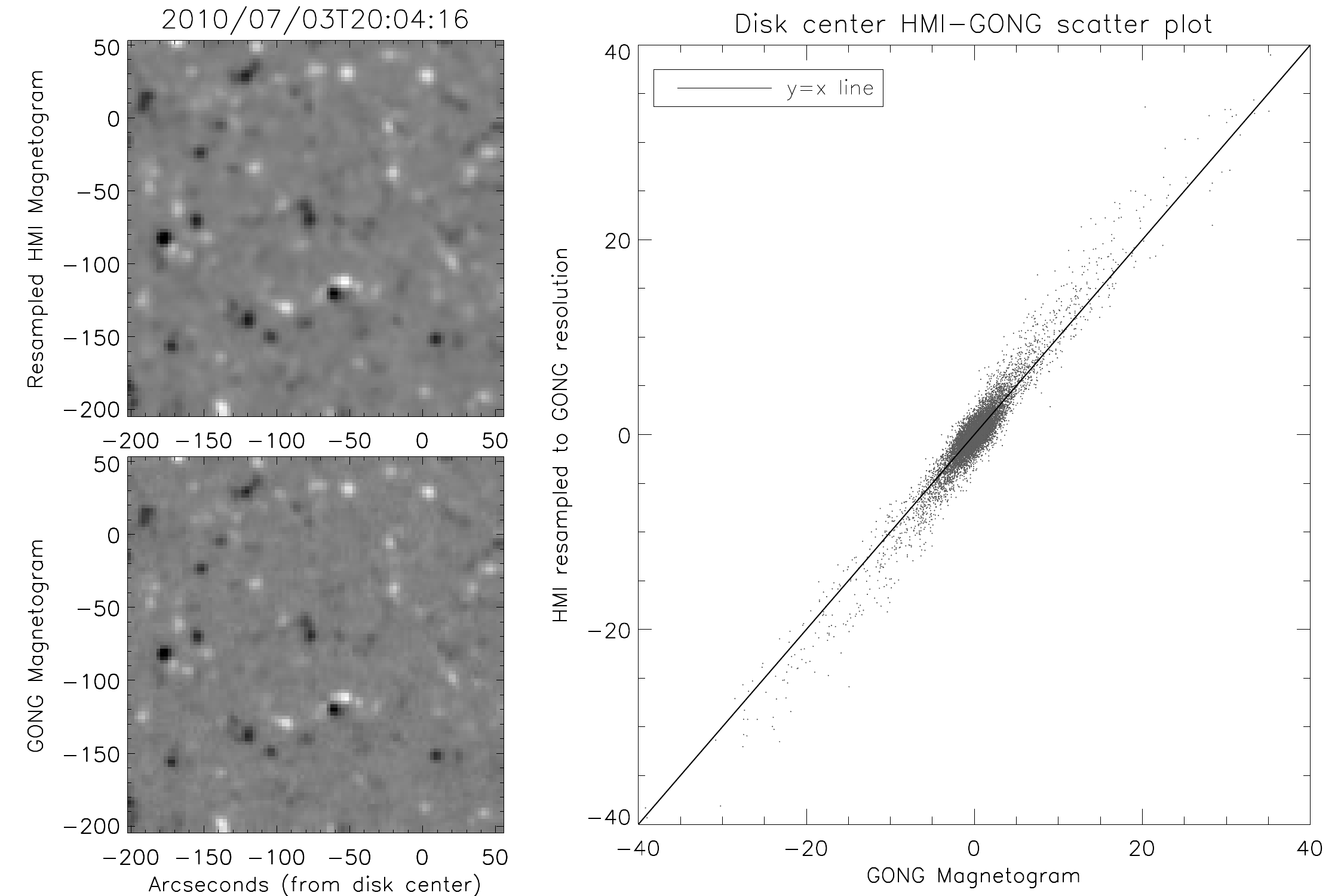}\end{center}
		\caption{Comparison between HMI (coaligned and resolution-matched) with GONG near disk center (highlighted area). There, GONG underestimates relative to HMI, whereas at the limb (Figure \ref{fig:hmi_gong_comparison_limb}), the reverse is true -- just as suggested by Table \ref{tab:calfactors}.}\label{fig:hmi_gong_comparison_center}
	\end{figure}

\begin{figure}
		\begin{center}\includegraphics[width=\textwidth]{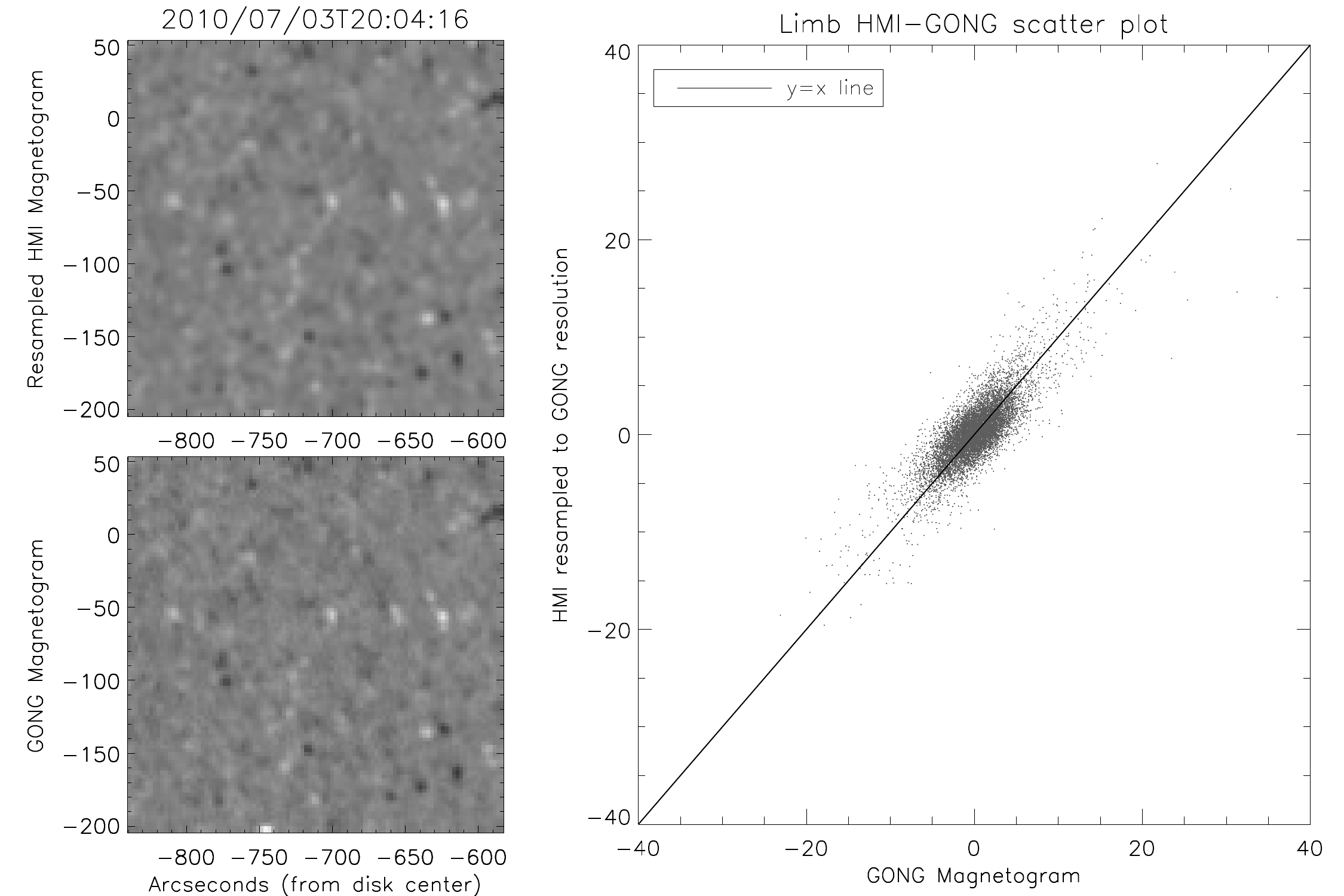}\end{center}
		\caption{Comparison between HMI (coaligned and resolution-matched) with GONG near the limb (highlighted area). There, GONG overestimates relative to HMI, whereas at disk center (Figure \ref{fig:hmi_gong_comparison_limb}, the reverse is true -- just as suggested by Table \ref{tab:calfactors}.}\label{fig:hmi_gong_comparison_limb}
	\end{figure}

	To check this behavior, we have compared HMI and GONG, being careful to match HMI's resolution with GONG's by convolving with the GONG PSF (the HMI PSF is much smaller than GONG's so its residual presence will have negligible effect), coalign, and match their pixel scale. This is presented in Figure \ref{fig:hmi_gong_comparison_center}, for disk center, and Figure \ref{fig:hmi_gong_comparison_limb} for the limb. Remarkably, very similar behavior to Table \ref{tab:calfactors} is observed: For disk center, GONG slightly underestimates relative to HMI, while on the limb they are very close, with GONG slightly overestimating relative to HMI. There is some spread in the scatterplots, it is true, but the relative correspondence with Table \ref{tab:calfactors} is rather striking.

	The sub-unity factors for 8x GONG resolution at 70 and 75 degrees are surprising and require some comment. In the case of the convective blueshift, that effect can reverse at high inclinations, becoming a convective redshift. It may be that the same thing is going on here, but in that case the question is why that doesn't happen for lower resolutions, especially since that effect should become less significant at higher resolution. A detailed investigation of this question is beyond the scope of the present work, but the following speculative explanations can be ventured:
	\begin{itemize}
		\item Although the other factors are not significantly sub-unity, they are still surprisingly small as well, and are likely also being affected by the transition from a blueshift-like effect to a redshift-like one.
		\item The improvement with resolution of a convective blueshift-like effects exhibits a thresholding behavior: If the resolution is enough to resolve the high-resolution bright/dark blueshifting/redshifting (or vice versa) structures, the effect vanishes. If the observations are not close to that resolution, on the other hand, there's little change in the behavior. Due to projection, the $ 75^\circ $ case has one quarter the resolution (in the direction perpendicular to the limb) of the disk center case, so the resolution of the 8x case may fall below the threshold on the limb but above it at disk center. The factor at disk center changes significantly between 4x and 8x GONG resolution, which suggests that 8x is close to that threshold but 1x and 2x are not.
		\item As mentioned above, it appears that additional effects are present which tend to cause underestimation of the field. These are likely to improve with resolution. If the 8x resolution case is not much affected by these effects but only by the (near limb) convective redshift-like effect, this would explain its overestimation of the field. 
	\end{itemize}

\begin{figure}
		\begin{center}\includegraphics[width=0.49\textwidth]{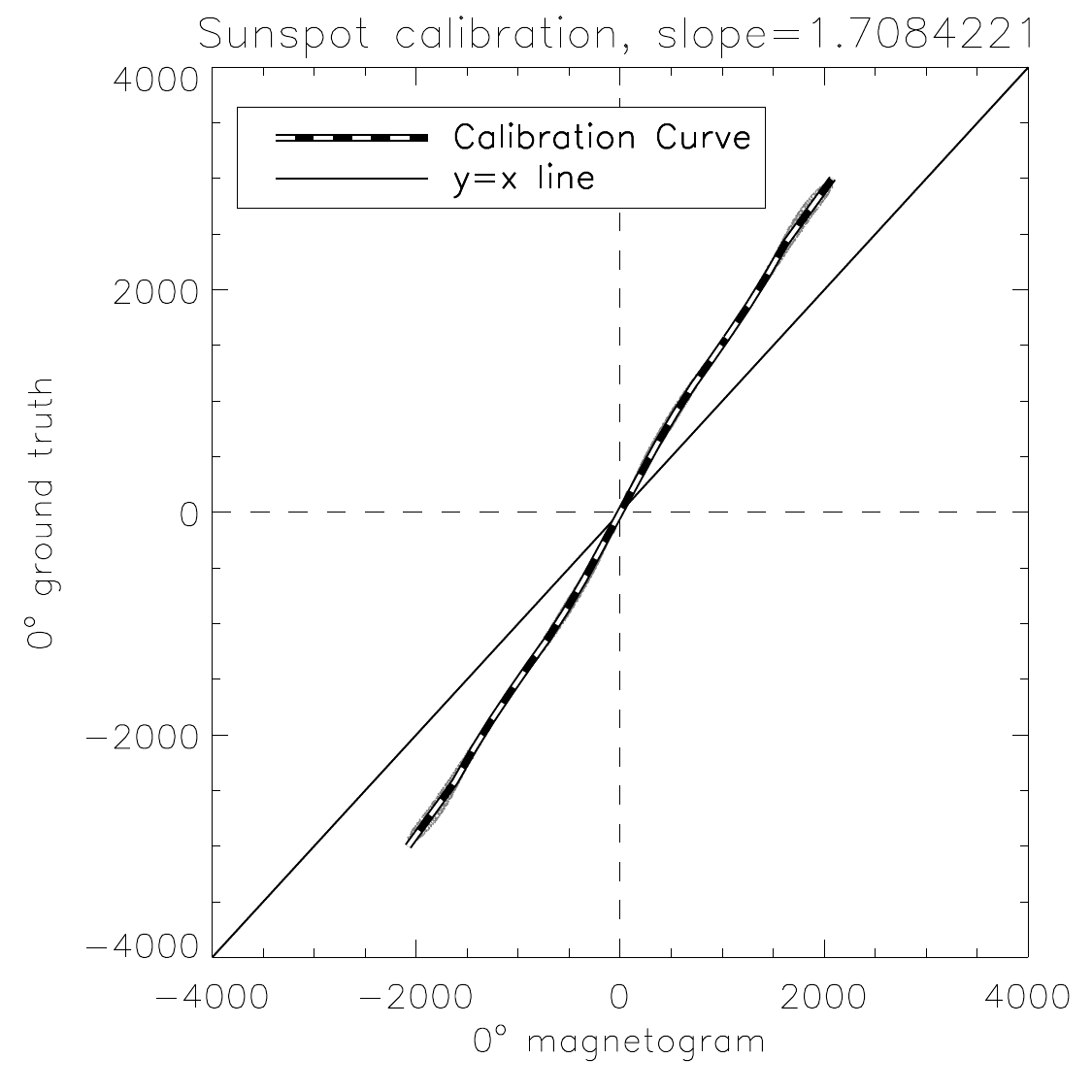}\includegraphics[width=0.49\textwidth]{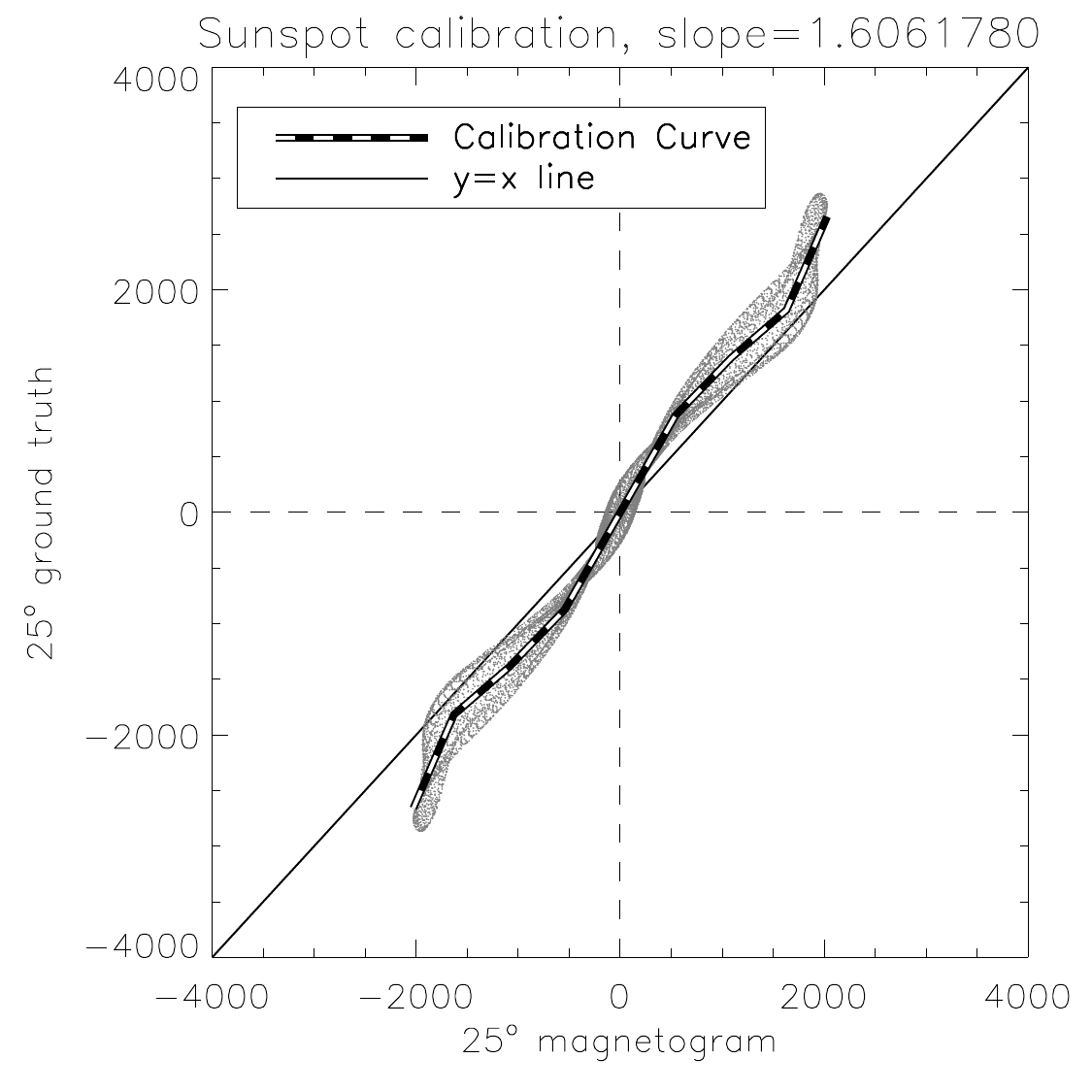}\end{center}
		\begin{center}\includegraphics[width=0.49\textwidth]{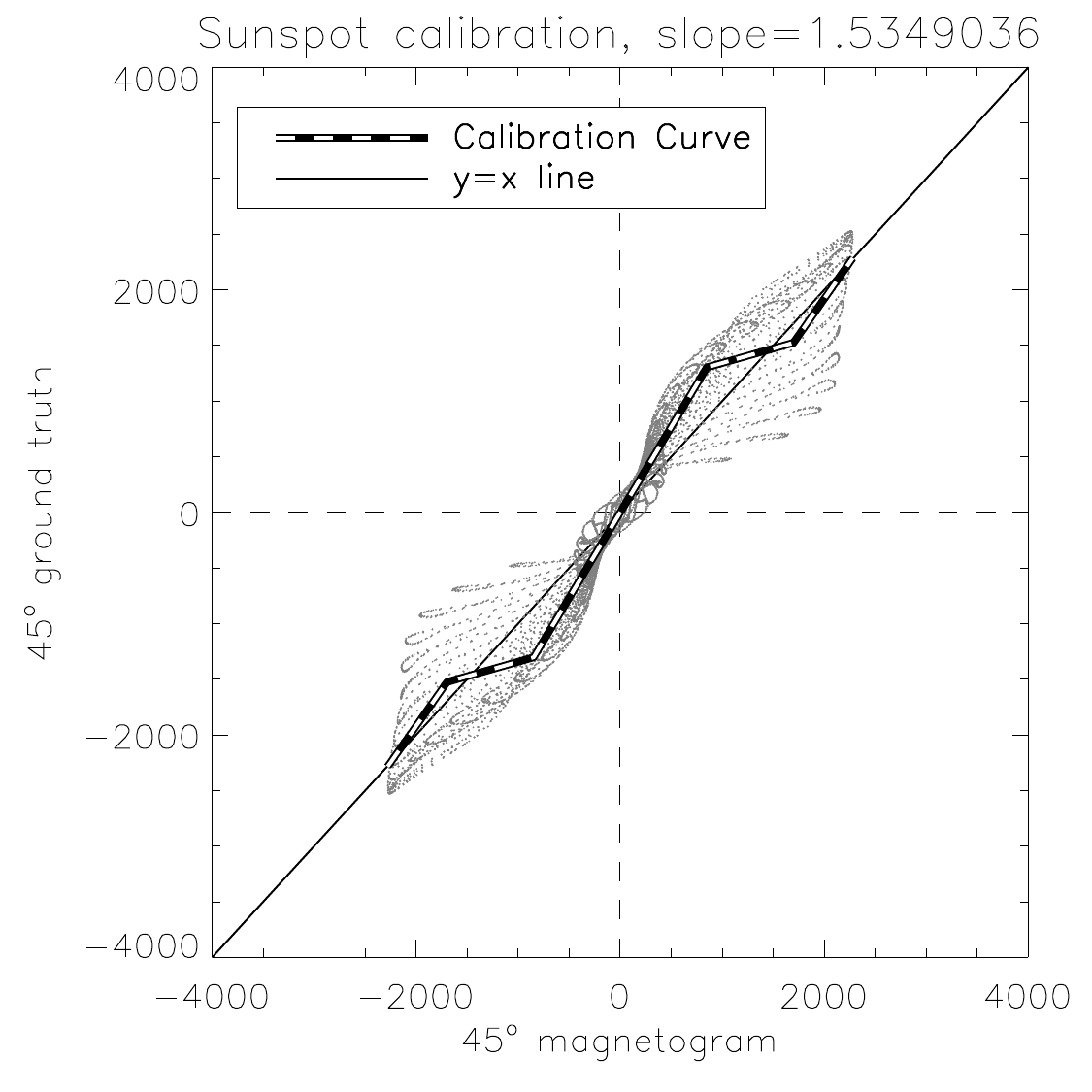}\includegraphics[width=0.49\textwidth]{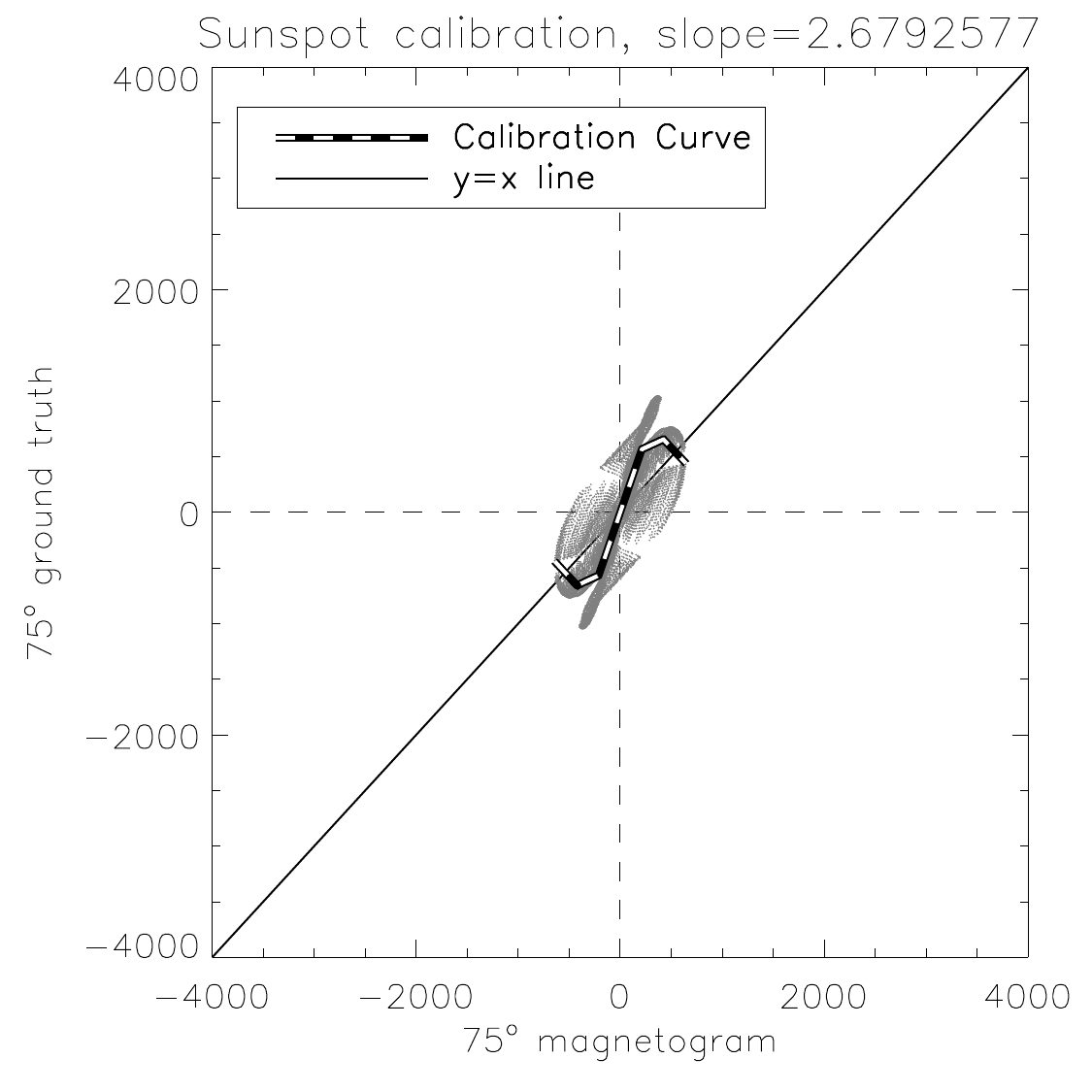}\end{center}
		\caption{Point clouds and calibration curves for sunspot at inclinations from 0 degrees (top left), 25 degrees (top right), 45 degrees (bottom left) and 75 degrees (bottom right). Small-field correction factors at low inclinations are $\sim 1.7$. Above 25 degree inclination, these point clouds are not usable for calibration purposes (see discussion in text).}\label{fig:sunspot_scatterplots}
	\end{figure}

	The preceding discussion has been of the `quiet sun', non-sunspot point clouds and calibration curves. These are the most important for space weather applications since they dominate the extrapolations \citep[e.g.,][]{Petrie2013}, but we also consider briefly the point clouds and corresponding curve for the sunspot. These are shown in Figure \ref{fig:sunspot_scatterplots}. Sunspot pixels are identified as areas which are dark in the pseudo-continuum intensity compared to their surroundings, and a fairly tight constraint is applied to ensure that quiet sun pixels do not contribute to the sunspot point clouds. At disk center, a very tight relationship with a slope of 1.77 is obtained. It appears that the sunspot correction factors are roughly constant between 0 and 45 degrees and then {\em increase} with latitude (opposite the behavior of the quiet sun case). 

	However, the quality of the point clouds quickly deteriorates with increasing inclination: even the one at $25^\circ $ is marginal, and all show pronounced features in the point clouds which are specific to this sunspot (the `tracks' seen in many of the point clouds correspond to specific features of the ground truth seen at differing subpixel offsets due to our tiling of the simulation). In the non-sunspot case, the point cloud represents a wide variety of small structures for which there is a typical ground truth field for any given measured field strength. That is not true for the sunspot: it is a monolithic structure and the higher inclination scatter plots make it clear that they have no `typical' ground truth field for a given measured field strength. Therefore, calibration by one-dimensional curve from a single sunspot is not sufficient at higher inclination. Multiple sunspot `ground truth' simulations and a more involved calibration method (e.g., neural network based) are likely to be necessary. Note also that we omit molecular lines from our radiative transfer, which is another limitation of sunspot calibration.

	This presents a dilemma for the present discussion. We are applying a significant correction factor to the non-sunspot fields, so if the magnetograms are to remain self-consistent the sunspots should have some correction as well. Fortunately, the low-inclination curves are the ones that are most important: sunspots are usually at low latitudes and the synoptic maps upon which most field extrapolations are based typically use only field values near the central meridian. As already mentioned, the sunspots are a secondary contributor to large-scale field extrapolations in any case \citep[][]{Petrie2013}. So the high-inclination sunspot calibration will not have a significant effect on data's primary use case. Since it appears that there are no issues with the disk center sunspot calibration curve, and those with the 25 degree sunspot curve are minor, we use only those sunspot curves in Section \ref{sec:calapplication}; higher inclination fields detected as sunspots have the 25 degree sunspot curve applied to them for visual continuity, but should be taken with a grain of salt. In particular, the trustworthyness of GONG sunspot measurements above 25 degrees inclination is unclear, especially for strong fields (see Figure \ref{fig:sunspot_scatterplots}), with calibration curve or without. Our treatment of these more inclined sunspot fields is for illustrative purposes and not meant to be prescriptive: a dedicated investigation of that question is indicated. 

	This primary issue with higher inclination sunspot fields may be due to the presence of the polarity inversion line. The abrupt change in slope of the curve at high field strengths suggests a difference in weighting of polarities in the measurement vs. that in the reduced ground truth. This can produce a drastic effect when the polarity reverses near where the field is strongest. That is exactly what happens at higher inclination, because the polarity inversion is near the middle of the sunspot. The synthetic magnetogram images (e.g., in Figures \ref{fig:example_scatterplot}, \ref{fig:cal_curve_25deg}, and \ref{fig:cal_curve_75deg}) illustrate this. This is exacerbated by GONG's low resolution: since the sunspot is only a few resolution elements tall at high inclination, the polarity inversion line contributes to a large fraction of the sunspot pixels. At high resolution the effect is reduced, since the fraction of pixels near enough to the polarity inversion line to be effected is much smaller. The sunspot point clouds at 2, 4, and 8 times GONG's resolution (not shown) are consistent with this, although their quality remains poor otherwise. 

	\subsection{Application of calibration curves}\label{sec:calapplication}

\begin{figure}
		\begin{center}\includegraphics[width=\textwidth]{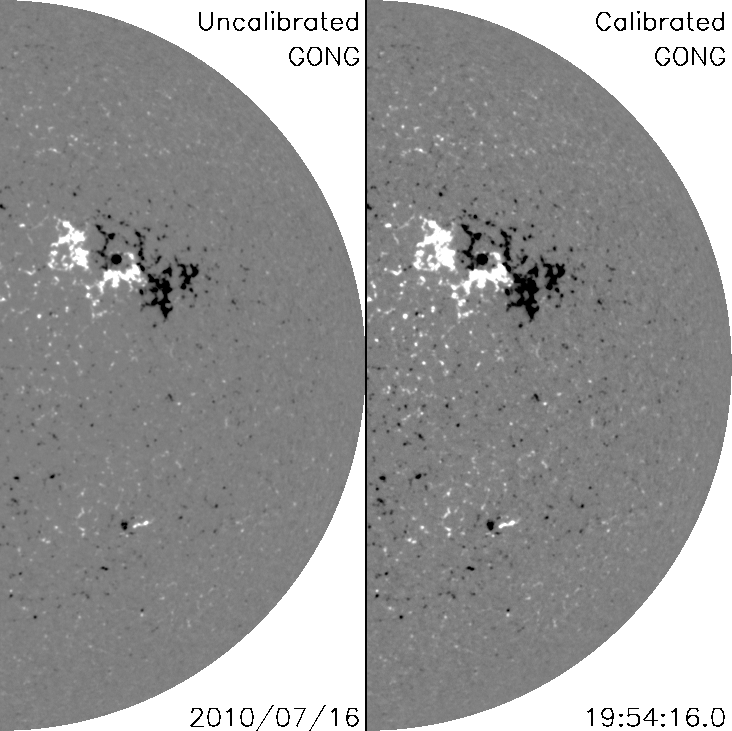}\end{center}
		\caption{Application of combined calibration curves to GONG magnetogram taken at 1954 UT on July 7, 2010. Left: original GONG magnetogram; Right: curve-corrected GONG magnetogram. Plotted magnetic field range is -100 to 100 Gauss.}\label{fig:example_calibrated_gong}
	\end{figure}

	To check the effects of these calibration curves, we have applied them to the period from June 8 to July 19, 2010. This is the same period investigated by \cite{LinkerOpenFlux2017}. In that paper, they find an open flux at 1 AU corresponding to $\sim 0.64$ nT average radial field strength based on GONG extrapolations, whereas the in situ observations (from OMNI) were $\sim 2$ nT. As a preliminary test of our calibration we performed a similar experiment for this time period using a PFSS extrapolation \citep[described in][]{Petrie2013}. As previously mentioned, the 0 and 25 degree sunspot curves are applied to sunspot regions (identified as those which are dark compared to their surroundings in the pseudo-continuum, with the same criteria as in the curve fitting); any sunspot regions at inclinations over 25 degrees use the 25 degree curve. Non sunspot regions use all 6 non-sunspot calibration curves. Figure \ref{fig:example_calibrated_gong} shows the calibrated magnetogram from one of these days (July 7) as an example, and compares it with the uncalibrated magnetogram.

\begin{figure}
		\begin{center}
			\includegraphics[width=0.49\textwidth]{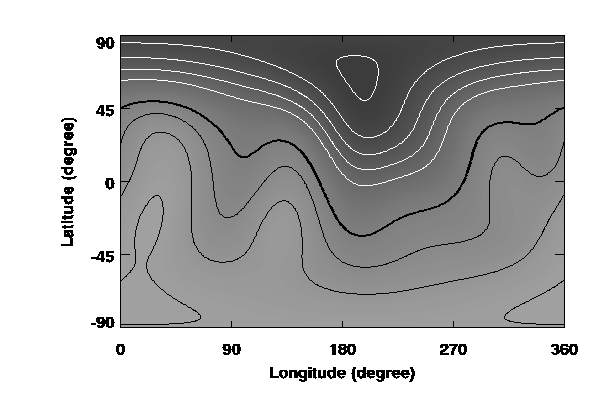}\includegraphics[width=0.49\textwidth]{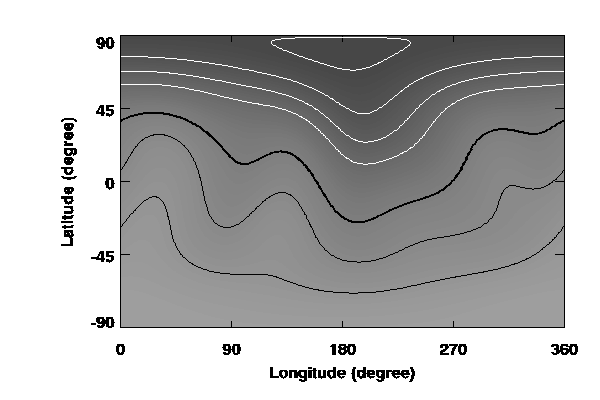}
		\end{center}
		\caption{Extrapolated flux distribution at the source surface resulting from preliminary application of our calibration curves to the June 8 to July 19, 2010 time period. Left shows the flux distribution with calibration, while right shows that without it. Heavy dark line in each shows location of the current sheet.}\label{fig:openflux_extrapolations}
	\end{figure}

	Our open flux comparison finds an average radial field strength at 1 AU of 0.42 nT without the calibration and 0.52 nT with it. The calibration therefore produces an increase in the open flux of $\sim 25\%$ with this fairly simple PFSS extrapolation. Our smaller values overall are likely due to the less sophisticated extrapolation, and the $25\%$ increase suggests the calibrated data might result in a GONG open flux of $\sim 0.8$ if the \cite{LinkerOpenFlux2017} analysis were applied to it. Figure \ref{fig:openflux_extrapolations} shows the flux distribution at the source surface resulting from our extrapolations. In addition to the overall stronger field, the calibrated extrapolation shows a more `distorted' field configuration, which would be relevant to space weather forecasting.

	\section{Conclusions, Discussion, and Summary}\label{sec:conclusions}


	These papers \citep{PlowmanEtal_2019I,PlowmanEtal_2019II,PlowmanEtal_2019III} have a number of significant findings, which we summarize here. \cite{PlowmanEtal_2019I} provides background, an overview of our analysis, a description of GONG and our simulation of it, and demonstrates that GONG is not subject to classical magnetograph saturation. Then, \cite{PlowmanEtal_2019II} examined the theory of calibration and magnetograph calibration, finding an issue that can arise with either one:

	\begin{itemize}
		\item Per-pixel comparison of magnetograms (e.g., scatter plots and histogram equating) will show apparent calibration differences unless their resolutions (including PSF) exactly match, typically appearing to show that a lower-resolution magnetograph's measurements are systematically lower than those of a higher-resolution magnetograph. The reason for this is that if the observations are dominated by unresolved and uncorrelated spatial variations (`salt \& pepper'), as they are on much of the sun, these fluctuations will cancel to a larger degree in the lower resolution instrument due to root $n$ averaging. 
		\item Per-pixel comparison of synthetic measurement and `reduced' ground truth will likewise show the same differences unless their resolutions (including PSF) exactly match. A calibration curve made from such a comparison will therefore tend to inflate the measurements, since the resolution of the ground truth is always at least as high as the synthetic measurements (usually it is much higher).
	\end{itemize}

	This difference is {\em not} indicative of a {\em real} calibration difference, however: it occurs even in the ideal case where the resolution differences do not add or remove flux from the magnetograms, only rearrange it. Calibration curves resulting in this case will therefore tend to inflate the flux. We demonstrated all of this for the ideal case by adding a PSF difference using a linear convolution -- the `linear' case. 

	This effect is therefore a `red herring' where magnetograph calibration or comparison is concerned: There is no need to account for it when using the magnetograms (e.g., for space weather extrapolation), only when making comparisons or (inter-)calibrations. To account for it in those cases, we advocated a the following solution, and demonstrated that it was effective (in the sense that the resulting calibration curves restored the ground truth fluxes to the synthetic measurements) in both the linear case and the more general case:

	\begin{itemize}
		\item Carefully match resolutions before making per pixel comparisons: For example, when comparing magnetograms, the higher resolution magnetogram should be reduced to the resolution (including PSF) of the lower resolution one. If the resolutions are not well characterized, both must be degraded to a significantly lower resolution than either PSF \citep[e.g.,][]{2010LambEtal_ApJ720_140,2013PietarilaEtal_SoPh282_91}, or they should be compared with a resolution-aware method \citep[e.g.,][]{VirtanenMursula2017}.
		\item Similarly, the resolution of the ground truth should be reduced to that of the synthetic measurements prior to making comparisons and calibration curves as we do in this work. 
	\end{itemize}
	
	These differences vanish when the magnetograms are compared at large spatial scales, and have no effect on the large-scale fluxes which are most important for space weather. This likely explains why, for example, \cite{Riley_comparison2014} finds significant differences in per pixel comparisons between GONG and HMI magnetograms, but when \cite{LinkerOpenFlux2017} compare extrapolations made with these instruments, the open fluxes are almost identical: The per-pixel flux comparisons of \cite{Riley_comparison2014} are different because the resolutions are not exactly matched, but in fact the fluxes measured by both instrument are very similar, leading to very similar extrapolated open fluxes for both. In this paper, we made a preliminary comparison of fluxes between HMI and GONG with resolution matching, and find that the fluxes are indeed very similar. Thus it appears likely that magnetograph measurements are in better agreement than the existing literature would indicate, and that a significant fraction of the reported apparent disagreement between magnetograms is caused by this effect.

	\noindent Some works in the literature \citep[e.g.,][]{2012LiuEtal_SoPh279_295L} have already employed a similar resolution-matching method, demonstrating that the need has already been recognized and that the solution has a peer reviewed track record. In this paper, we have employed this approach when constructing our calibration curves from the full GONG simulator results. We now turn to those results:
	
	As with the preliminary results in \cite{PlowmanEtal_2019II}, we find that there is also a {\em real} effect which casuses synthetic GONG measurements underestimate their fluxes by factor of over 2 compared to the flux {\em over the same area} in the MURaM ground truth. In \cite{PlowmanEtal_2019II}, we explained that this effect is likely similar to the `convective blueshift': the unresolved granulation pattern biases the measurements to the brighter regions (the granule centers), which have weak fields. This is for the `quiet sun' (non-sunspot) regions at disk center; the factor drops to near 1 at the limb, which is also consistent with the convective blueshift. We investigated the effects of resolution and found that it persists as long as the resolution is too low to resolve the granulation pattern. The situation with the sunspot calibration is more complex; those results are summarized as follows:
	

	\begin{itemize}
		\item A clear calibration curve relationship is found at disk center. For weak fields, the slope is $\sim 1.73$, and drops to $\sim 1.5$ for stronger fields. Thus the miscalibration at disk center is somewhat less for sunspots than for quiet sun (non-sunspots).
		\item Unlike the non-sunspot case, the degree of underestimation appears to {\em increase} at large inclinations.
		\item The quality of the scatter plots deteriorates rapidly with inclination, such that above 45 degrees there is no clear `typical' ground truth field for a given measured field strength at 45 degrees or above. Thus, the point cloud and curve approach is unsuitable to calibration of sunspot fields above $\sim 25$ degrees. 
		\item For the same reason, sunspot magnetograms at high inclination may not be trustworthy: there is not a clear monotonic relationship between the measured field strength at the ground truth field that produced it. Caution should be exercised in the use and interpretation of these measurements.
		\item Because this effect is largely absent at zero inclination, concurrent observations from Earth and a vantage point away from the Earth-sun line are likely to prove invaluable in understanding and correcting for these effects.
	\end{itemize}

	These results suggest that all synoptic magnetograms are, to some extent, underestimating the real solar magnetic fluxes, except at high inclination in the quiet sun. Therefore they will all tend to underestimate the extrapolated fields, as has been found in the literature, although a preliminary check of the open flux using this new calibration finds a factor of 25\% increase, overall, in the open flux. We have therefore accounted for a significant fraction of the $\sim 2$ factor needed to make extrapolations match the {\em in situ} flux observations.

	To summarize, our results indicate that the large apparent disagreements found in some previous comparisons of magnetographs do not reflect a real difference in their relative flux calibrations and are therefore a red herring. The open flux discrepancies, on the other hand, are due in part to a real effect that causes magnetographs to underestimate their fluxes, by more than a factor of two in some cases. As expected, magnetograph calibration does not appear to be able to account for all of the missing flux alone, and the other usual suspects are still in play: longitudinal-to-radial conversion, the source surface assumption, linear vs. nonlinear field extrapolations in general, etc.

	Finally, we point out that observations from NSO's new Daniel K. Inouye Solar Telescope (DKIST) will resolve the granulation pattern, and should be unaffected by this convective blueshift-like effect. As a result, we predict that DKIST Zeeman effect measurements (e.g., from ViSP) of the magnetic flux at disk center will be a factor of $\sim 2$ higher than HMI (or GONG) measurements of the same fluxes (making sure to take the much larger integration area of HMI or GONG into account by integrating the DKIST fluxes over those areas). DKIST measurements will also be very interesting for investigating these effects at high latitude.

	These papers have been the first phase of a research project funded under a NASA space weather `operations to research' (O2R) grant whose specific goal is to improve the accuracy of solar wind prediction models. In subsequent phases we address polar field measurements in more detail, as well as apply the `re-calibrated' GONG magnetograms to WSA/Enlil model runs for known solar wind and CME events that were particularly poorly forecast.

	\subsection*{Acknowledgements}
	This work was funded in part by the NASA Heliophysics Space Weather Operations-to-Research program, grant number 80NSSC19K0005, and by a University of Colorado at Boulder Chancellor’s Office Grand Challenge grant for the Space Weather Technology, Research, and Education Center (SWx TREC).

	We acknowledge contributions, discussion, information, and insight from a variety of sources: Gordon Petrie, Jack Harvey, Valentin Martinez Pillet, Sanjay Gosain, and Frank Hill, among others.

	\section*{Disclosure of Potential Conflicts of Interest}
	The authors declare that they have no conflicts of interest.

	\bibliographystyle{apj}
	\bibliography{apj-jour,GONG_simulatorIII}
\end{document}